\documentstyle{cupconf}

\ifoldfss
\else
  \ifnfssone
    \newmathalphabet{\mathit}
      \addtoversion{normal}{\mathit}{cmr}{m}{it}
      \addtoversion{bold}{\mathit}{cmr}{bx}{it}
    \newmathalphabet{\mathcal}
      \addtoversion{normal}{\mathcal}{cmsy}{m}{n}
    \else
    \ifnfsstwo
    \fi
  \fi
\fi

\def\hexnumber#1{\ifcase#1 0\or1\or2\or3\or4\or5\or6\or7\or8\or9\or
 A\or B\or C\or D\or E\or F\fi }

\makeatletter
\ifx\CUP@mtlplain@loaded\undefined
\else
\fi
\makeatother
 \makeatletter
 \ifx\CUP@mtlplain@loaded\undefined
   \font\tenbmi=cmmib10 at 10pt
   \font\sevenbmi=cmmib10 at 7pt
   \font\fivebmi=cmmib10 at 5pt

   \newfam\bmifam
   \textfont\bmifam=\tenbmi
   \scriptfont\bmifam=\sevenbmi
   \scriptscriptfont\bmifam=\fivebmi
   
 \fi
 \makeatother
\ifnfsstwo

\fi
\ifnfssone

\fi
\ifoldfss

\fi

\mathchardef\varLambda="0103

\makeatletter
\ifx\CUP@mtlplain@loaded\undefined
\else
\fi
\makeatother
\makeatletter
\ifx\CUP@mtlplain@loaded\undefined
  \font\tenbms=cmbsy10
  \font\sevenbms=cmbsy10 at 7pt
  \font\fivebms=cmbsy10 at 5pt
  \newfam\bmsfam
  \textfont\bmsfam=\tenbms
  \scriptfont\bmsfam=\sevenbms
  \scriptscriptfont\bmsfam=\fivebms

  \edef\bsy@{\hexnumber\bmsfam}
  \mathchardef\bnabla="0\bsy@72
\fi
\makeatother
\newcommand{\beq}{\begin{equation}}
\newcommand{\eeq}{\end{equation}}
\newcommand{\eprime}{\epsilon^{\prime}}
\catcode`\@=11 
\def\lsim{\mathrel{\mathpalette\@versim<}}
\def\gsim{\mathrel{\mathpalette\@versim>}}
\def\@versim#1#2{\vcenter{\offinterlineskip
        \ialign{$\m@th#1\hfil##\hfil$\crcr#2\crcr\sim\crcr } }}
\catcode`\@=12 
\input{psfig}
\newcommand{\apj}{{\em The Astrophysical Journal}}
\newcommand{\apjl}{{\em The Astrophysical Journal Letters}}
\newcommand{\araa}{{\em Annual Reviews of Astronomy \& Astrophysics}}
\newcommand{\aanda}{{\em Astronomy \& Astrophysics}}
\newcommand{\aandas}{{\em Astronomy \& Astrophysics Supplement}}
\newcommand{\mnras}{{\em Monthly Notices of the Royal Astronomical Society}}
\newcommand{\pasj}{{\em Publications of the Astronomical Society of Japan}}
\newcommand{\prl}{{\em Physical Review Letter}}
\newcommand{\nature}{{\em Nature}}

\begin{document}
\ifnfssone
\else
  \ifnfsstwo
  \else
    \ifoldfss
      \let\mathcal\cal
      \let\mathrm\rm
      \let\mathsf\sf
    \fi
  \fi
\fi

  \title[Advection-Dominated Accretion]{Advection-Dominated Accretion around Black Holes\footnote{To appear in {\em The Theory of Black Hole Accretion
Discs}, eds. M. A. Abramowicz, G. Bjornsson, \& J. E. Pringle (Cambridge 
University Press)}}

  \author[Narayan, Mahadevan \& Quataert]{
  R\ls A\ls M\ls E\ls S\ls H\ns
  N\ls A\ls R\ls A\ls Y\ls A\ls N\ns \\
  R\ls O\ls H\ls A\ls N\ns  M\ls A\ls H\ls A\ls D\ls E\ls V\ls A\ls N\ns \\
	\and  \\
  E\ls L\ls I\ls O\ls T\ns Q\ls U\ls A\ls T\ls A\ls E\ls R\ls T\ls
  }
  \affiliation{Harvard--Smithsonian Center for Astrophysics,
    60 Garden Street, Cambridge, MA 02138, USA\\[\affilskip]
    }
  \maketitle

\begin{abstract}
This article reviews the physics of advection-dominated accretion
flows (ADAFs) and describes applications to several black hole X-ray
binaries and galactic nuclei.  The possibility of using ADAFs to
explore the event horizons of black holes is highlighted.
\end{abstract}

\firstsection 
\section{Introduction} 

Accretion processes around black holes almost inevitably involve
rotating gas flows.  Consequently, there is great interest in
self--consistent solutions of the hydrodynamic equations of viscous
differentially--rotating flows.  Four solutions are currently known
(see Chen et al. 1995 for a discussion).  In these solutions viscosity
transports angular momentum outward, allowing the accreting gas to
spiral in toward the central mass.  Viscosity also acts as a source of
heat; some or all of this heat is radiated, leading to the
observed spectrum.

The most famous of the four solutions is the thin disk model developed
by Shakura \& Sunyaev (1973), \cite{nt73}, \cite{lp74} and others (see
Pringle 1981 and Frank et al. 1992 for reviews).  The accreting gas
forms a geometrically thin, optically thick disk, and produces a
quasi--blackbody spectrum.  The effective temperature of the radiation
is in the range $10^5 - 10^7$ K, depending on the black hole mass and
the accretion rate ($T_{\rm eff} \propto M^{-1/4} \dot{M}^{1/4}$).
The thin disk solution has been used to model a large number of
astrophysical systems.

Shapiro, Lightman and Eardley (1976; hereafter SLE; see also
Bj\"ornsson \& Svensson 1991 and Luo \& Liang 1994) discovered a
second, much hotter self--consistent solution in which the accreting
gas forms a two temperature plasma with the ion temperature greater
than the electron temperature ($T_i \sim 10^{11}$K, $T_e \sim
10^8-10^{9}$K).  The gas is optically thin and produces a power--law
spectrum at X--ray and soft $\gamma$--ray energies.  The SLE solution
is, however, thermally unstable (Piran 1978), and is therefore not
considered viable for real flows.

At super--Eddington accretion rates, a third solution is present (Katz
1977; Begelman 1978; Abramowicz et al. 1988; see also Begelman \&
Meier 1982; Eggum, Coroniti, \& Katz 1988), in which the large optical
depth of the inflowing gas traps most of the radiation and carries it
inward, or ``advects'' it, into the central black hole.  This solution
is referred to as an optically thick advection--dominated accretion
flow (optically thick ADAF).  A full analysis of the dynamics of the
solution was presented in an important paper by \cite{Ab_et_al88}.

A fourth solution is present in the opposite limit of low,
sub--Eddington, accretion rates (Ichimaru 1977; Rees et al. 1982;
Narayan \& Yi 1994, 1995a, 1995b; Abramowicz et al. 1995).  In this
solution, the accreting gas has a very low density and is unable to
cool efficiently within an accretion time. The viscous energy is
therefore stored in the gas as thermal energy instead of being
radiated, and is advected onto the central star.  The gas is optically
thin and adopts a two--temperature configuration, as in the SLE
solution. The solution is therefore referred to as an optically thin
ADAF or a two--temperature ADAF.

This article reviews ADAFs around black holes, with an emphasis on
two--temperature ADAFs.

\section{Dynamics of ADAFs}

\subsection{Basic Equations}

Consider a steady axisymmetric accretion flow.  The dynamics of the
flow are described by the following four height-integrated
differential equations, which express the conservation of mass, radial
momentum, angular momentum, and energy: \beq {d \over dR}(\rho R H v)
= 0,
\label{masscons} \eeq \beq v {dv \over dR} - \Omega^2 R = -
\Omega_K^2R - {1 \over \rho} \, {d \over dR } ( \rho c_s^2),
\label{radialeq} \eeq \beq v \, {d(\Omega R^2) \over dR} = {1 \over
\rho R H} \, {d \over dR} \left( \nu \rho R^3 H {d\Omega \over dR}
\right), \label{angulareq} \eeq \beq \rho v T {ds\over dR} = q^+ - q^- = \rho
\nu R^2 \left( {d\Omega \over dR} \right)^2 - q^- \equiv f
\nu \rho R^2 \left({d \Omega \over dR}\right)^2,
\label{energyeq}
\eeq where $\rho$ is the density of the gas, $R$ is the radius, $H\sim
c_s/\Omega_K$ is the vertical scale height, $v$ is the radial
velocity, $c_s$ is the isothermal sound speed, $T$ is the temperature
of the gas (mean temperature in the case of a two temperature gas),
$\Omega$ is the angular velocity, $\Omega_K$ is the Keplerian angular
velocity, $s$ is the specific entropy of the gas, $q^+$ is the energy
generated by viscosity per unit volume, and $q^-$ is the radiative
cooling per unit volume.  The quantity on the left in equation
(\ref{energyeq}) is the rate of advection of energy per unit
volume. The parameter $f$ is thus the ratio of the advected energy to
the heat generated and measures the degree to which the flow is
advection--dominated.  The kinematic viscosity coefficient, $\nu$, is
generally parameterized via the $\alpha$ prescription of Shakura \&
Sunyaev (1973), \beq \nu \equiv \alpha c_s H = \alpha {c_s^2 \over
\Omega_K}, \eeq where $\alpha$ is assumed to be independent of $R$.
The steady state mass conservation equation implies a constant
accretion rate throughout the flow: \beq \dot{M} = (2\pi R)(2H)\rho
|v|=\mbox{constant}.  \eeq

It is useful to rewrite the energy equation (\ref{energyeq}) compactly
as \beq q^{\rm adv} = q^+ - q^-, \eeq where $q^{\rm adv}$ represents
the advective transport of energy (usually a form of cooling).
Depending on the relative magnitudes of the terms in this equation,
three regimes of accretion may be identified:
\begin{itemize}
\item $q^+ \simeq q^- \gg q^{\rm adv}$.  This corresponds to a
cooling-dominated flow where all the energy released by viscous
stresses is radiated; the amount of energy advected is negligible. The
thin disk solution and the SLE solution correspond to this regime.
\item $q^{\rm adv} \simeq q^+ \gg q^-$.  This corresponds to an ADAF
where almost all the viscous energy is stored in the gas and is
deposited into the black hole.  The amount of cooling is negligible
compared with the heating.  For a given $\dot{M}$, an ADAF is much
less luminous than a cooling-dominated flow.
\item $-q^{\rm adv} \simeq q^- \gg q^+$.  This corresponds to a flow
where energy generation is negligible, but the entropy of the
inflowing gas is converted to radiation. Examples are Bondi accretion,
Kelvin--Helmholtz contraction during the formation of a star, and
cooling flows in galaxy clusters.
\end{itemize}

\subsection{Self--Similar Solution}
\label{secselfsim}

Analytical approximations to the structure of optically thick and
optically thin ADAFs have been derived by Narayan \& Yi (1994).
Assuming Newtonian gravity ($\Omega_K^2 = GM/R^3$) and taking $f$ to
be independent of $R$, they showed that equations
(\ref{masscons})--(\ref{energyeq}) have the following self--similar
solution (see also Spruit et al. 1987):
\beq
v(R) = -{(5+2\eprime)  \over 3 \alpha^2} g(\alpha, \eprime) \alpha v_{\rm ff},
\label{soln6}
\eeq
\beq
\Omega(R) = \left[{2\eprime(5+2\eprime) \over 9\alpha^2} g(\alpha, \eprime)
\right]^{1/2} \, {v_{\rm ff}\over R},
\eeq
\beq
c_s^2(R) = {2(5+2\eprime) \over 9\alpha^2} \, g(\alpha, \eprime)
v_{\rm ff},
\label{soln8}
\eeq where \beq v_{\rm ff} \equiv \left( {GM \over R}\right)^{1/2}, \
\ \ \eprime \equiv {\epsilon \over f} = {1 \over f} \left( {5/3 -
\gamma \over \gamma -1} \right), \ \ \ g(\alpha, \eprime) \equiv \left
[ 1 + {18 \alpha^2 \over (5+2\eprime)^2} \right]^{1/2} - 1.  \eeq
$\gamma$ is the ratio of specific heats of the gas, which is likely to
lie in the range 4/3 to 5/3 (the two limits correspond to a radiation
pressure--dominated and a gas pressure--dominated accretion flow,
respectively).  Correspondingly, $\epsilon$ lies in the range 0 to 1.

In general, $f$ depends on the details of the heating and cooling and
will vary with $R$. The assumption of a constant $f$ is therefore an
oversimplification. However, when the flow is highly advection
dominated, $f\sim 1$ throughout the flow, and $f$ can be well
approximated as constant.  Setting $f=1$ and taking the limit
$\alpha^2 \ll 1$ (which is nearly always true), the solution
(\ref{soln6})--(\ref{soln8}) takes the simple form \beq {v \over
v_{\rm ff}} \simeq -\left({\gamma - 1 \over \gamma -5/9}\right)
\alpha, \ \ {\Omega \over \Omega_K} \simeq \left[{2(5/3 -\gamma) \over
3(\gamma -5/9)}\right]^{1/2}, \ \ {c_s^2 \over v^2_{\rm ff}} \simeq
{2\over 3} \,\left( {\gamma - 1 \over \gamma-5/9}\right).
\label{simpleeq} \eeq

A number of interesting features of ADAFs are revealed by the
self--similar solution.  (1) As we discuss below (\S 4.1.3), it
appears that ADAFs have relatively large values of the viscosity
parameter, $\alpha \gsim 0.1$; typically, $\alpha \sim 0.2-0.3$.  This
means that the radial velocity of the gas in an ADAF is comparable to
the free--fall velocity, $v \gsim 0.1 v_{\rm ff}$.  The gas thus
accretes quite rapidly. (2) The gas rotates with a sub--Keplerian
angular velocity and is only partially supported by centrifugal
forces.  The rest of the support is from a radial pressure gradient,
$\nabla P\sim \rho c_s^2 /R\sim (0.3-0.4) v_{\rm ff}^2 /R $.  In the
extreme case when $\gamma \rightarrow 5/3$, the flow has no rotation
at all ($\Omega \rightarrow 0$).  (3) Since most of the viscously
generated energy is stored in the gas as internal energy, rather than
being radiated, the gas temperature is quite high; in fact, optically
thin ADAFs have almost virial temperatures.  This causes the gas to
``puff up'': $H\sim c_s/\Omega_K \sim v_{\rm ff}/\Omega_K \sim R$.
Therefore, geometrically, ADAFs resemble spherical Bondi (1952)
accretion more than thin disk accretion.  It is, however, important to
note that the {\em dynamics} of ADAFs are very different from that of
Bondi accretion (\S2.7). (4) The gas flow in an ADAF has a positive
Bernoulli parameter (Narayan \& Yi 1994, 1995a), which means that if
the gas were somehow to reverse its direction, it could reach infinity
with net positive energy.  This suggests a possible connection between
ADAFs and jets. (5) The entropy of the gas increases with decreasing
radius.  ADAFs are therefore convectively unstable (Narayan \& Yi
1994, 1995a; Igumenshchev, Chen, \& Abramowicz 1996).

\subsection{Vertical Structure}

The quasi--spherical nature of the gas flow in an ADAF might indicate
that the use of height integrated equations is an
oversimplification. This led Narayan \& Yi (1995a) to investigate the
structure of ADAFs in the polar direction, $\theta$.  Making use of
non--height--integrated equations, they searched for radially
self--similar solutions; for example, they assumed that the density
scales as $\rho \propto R^{-3/2} \rho(\theta)$, with a dimensionless
function $\rho(\theta)$; the radial velocity scales as $v \propto
R^{-1/2} v(\theta)$, etc..  The results confirmed that ADAFs are
quasi--spherical and quite unlike thin disks.

\begin{figure}
\centerline{
\psfig{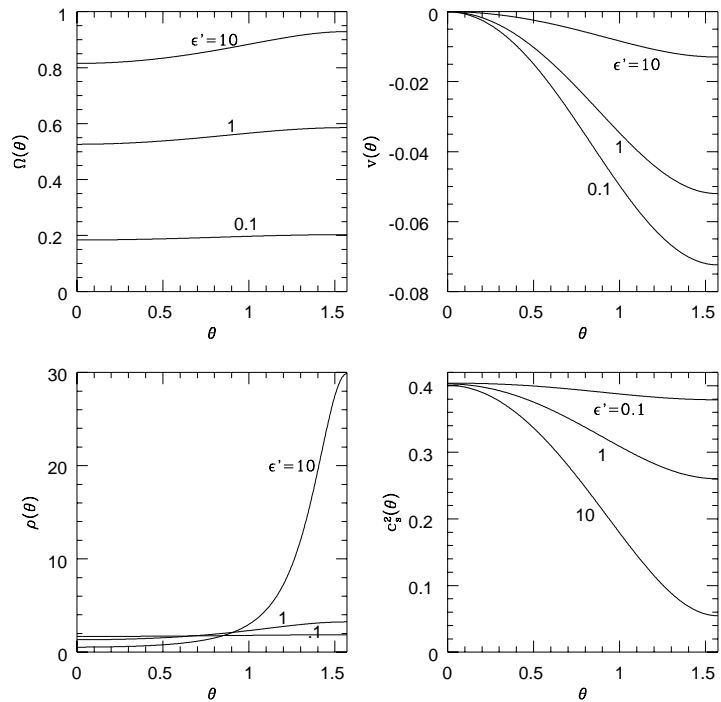}
}
\caption{Angular profiles for radial self-similar solutions
with $\alpha = 0.1$, $\eprime = 0.1, 1, 10$.  {\em Top
left:} angular velocity $\Omega/\Omega_K$ as a function of polar angle
$\theta$.  {\em Top right:} radial velocity, $v/v_{\rm ff}$.  {\em Bottom
left:} density, $\rho$.  {\em Bottom right:} sound speed squared,
$c^2_s/v^2_{\rm ff}$.}
\label{isodensity}
\end{figure}

Figure 1 shows the variation of $\Omega, v, \rho$ and $c_s^2$ as a
function of $\theta$ for some typical solutions ($\alpha = 0.1,
\eprime = 0.1, 1, 10$). For fully advection dominated flows ($f=1$),
different values of $\eprime$ correspond to different values of
$\gamma$: the three solutions shown have $\gamma = 1.61, 1.33, 1.06$.
If $\gamma$ is fixed, than increasing $\eprime$ corresponds to
decreasing $f$.  Large values of $\eprime$ therefore correspond to
cooling--dominated thin disk solutions whereas small values of
$\eprime$ correspond to highly advection--dominated flows.  From
Figure \ref{isodensity}, we find that $\Omega, \rho$ and $c_s^2$ in
ADAFs ($\eprime = 0.1, 1$) are nearly constant on radial shells.  The
radial velocity $v$ is zero at the poles and reaches a maximum at the
equator.  Most of the accretion therefore takes place in the
equatorial plane.  The figure also shows the expected features of a
thin disk in the limit of efficient cooling ($\eprime = 10$); the
density is peaked near the equator, and $\Omega$ approaches
$\Omega_K$.

Despite the quasi--spherical nature of the flow, Narayan \& Yi (1995a)
found that the solutions of the exact non--height--integrated
equations agree quite well with those of the simplified
height--integrated equations (\S \ref{secselfsim}), provided that
``height--integration'' is done along $\theta$ at constant spherical
radius, rather than along $z$ at constant cylindrical radius.  The
height--integrated equations therefore are a fairly accurate
representation of quasi--spherical ADAFs. (Technically, this has been
proved only in the self-similar regime.)

\subsection{Pseudo--Newtonian Global Solutions}

The self--similar solution is scale free and does not match the
boundary conditions of the flow.  To proceed beyond self-similarity it
is necessary to solve the full equations
(\ref{masscons})--(\ref{energyeq}) with proper boundary conditions.
For example, far from a central black hole, an ADAF might join on to a
thin disk.  At this radius, the rotational and radial velocities, as
well as the gas density and sound speed, must take on values
appropriate to a thin disk.  Also, as the accreting gas flows in
toward the black hole, it must undergo a sonic transition at some
radius $R_{sonic}$, where the radial velocity equals the local sound
speed.  In addition, since the black hole cannot support a shear
stress, the torque at the horizon must be zero, which is yet another
boundary condition to be satisfied by the solutions. (Alternatively,
if a causal viscosity prescription is used, there is a boundary
condition at the causal horizon; see, e.g., Gammie \& Popham (1998))
From the simple scalings of the self--similar solution
(\ref{soln6})--(\ref{soln8}), it is evident that the solution does not
satisfy the boundary conditions and, therefore, is not a good
approximation near the flow boundaries.  In fact, the question arises:
is the self--similar solution a good approximation to the global flow
at any radius?

This question was investigated by Narayan, Kato, \& Honma (1997a) and
Chen, Abramowicz \& Lasota (1997; see also Matsumoto et al. 1985;
Abramowicz et al. 1988; Narayan \& Yi 1994).  Integrating the angular
momentum equation (eq. [\ref{angulareq}]) once, one obtains \beq
{d\Omega \over dR} = {v\Omega_K(\Omega R^2 - j) \over \alpha R^2
c_s^2},\label{eigenval} \eeq where the integration constant $j$ is the
angular momentum per unit mass accreted by the central mass. Recalling
that eq. (\ref{masscons}) integrates to give the constant mass
accretion rate, $\dot{M}$, the global steady state problem consists of
solving the differential equations (\ref{radialeq}), (\ref{energyeq}),
and (\ref{eigenval}), along with proper boundary conditions, to obtain
$v(R), \ \Omega(R), \ c_s(R), \ \ln[\rho(R)]$, and the eigenvalue $j$.

Narayan et al. (1997a) and Chen et al. (1997) obtained global
solutions for accretion on to a black hole, using a pseudo--Newtonian
potential (Paczy\'nski \& Wiita 1980), with $\phi(R) = -GM/(R-R_S)$
and $\Omega_K^2 = GM/(R-R_S)^2 R$, which simulates a Schwarzschild
black hole of radius $R_S= 2GM/c^2$.  The global solutions agree quite
well with the self-similar solutions, except near the boundaries,
where there are significant deviations.  This means that the
self-similar solution provides a good approximation to the real
solution over most of the flow.

\begin{figure}
\centerline{
\psfig{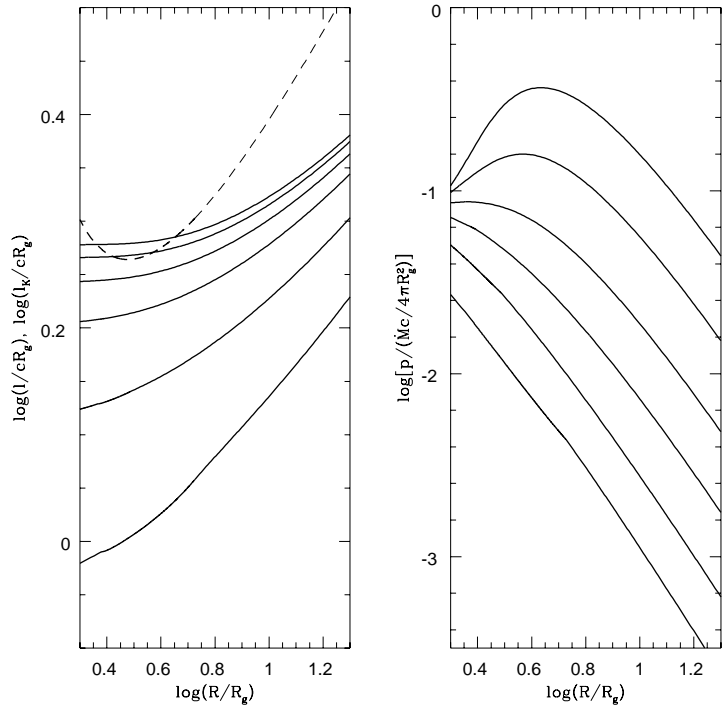}
}
\caption{ {\em left}: Radial variation of the specific angular
momentum, $l$, for (solid lines, from top to bottom) $\alpha$ = 0.001,
0.003, 0.01, 0.03, 0.1, and 0.3.  The dashed line shows the Keplerian
specific angular momentum, $l_K$.  Note that the low $\alpha$
solutions (0.001 and 0.003) are super-Keplerian ($l > l_K$) over a
range of radii, whereas the solutions with larger $\alpha$ have $l <
l_K$ for all radii.  {\em right}: Radial variation of the gas pressure
for the same $\alpha$.  Note that the low $\alpha$ curves (the two
upper ones) have pressure maxima, associated with the super-Keplerian
rotation shown in the left panel (taken from Narayan et al. 1997a).}
\label{alpha}
\end{figure}

Figure \ref{alpha} shows the radial variation of the specific angular
momentum and gas pressure in the inner regions of global ADAF
solutions for several $\alpha$.  It is seen that low $\alpha$ ADAFs
are quite different from high $\alpha$ ADAFs, the division occurring
at roughly $\alpha \sim 0.01$. When $\alpha \lsim 0.01$, the gas in an
ADAF is somewhat inefficient at transferring angular momentum
outwards, has low radial velocity, and is nearly in hydrostatic
equilibrium.  The flow is super--Keplerian over a range of radii and
has an inner edge near the marginally stable orbit (see
fig. \ref{alpha}).  The super--Keplerian flow creates a funnel along
the rotation axis, which leads to a toroidal morphology.  These
solutions closely resemble the thick--torus models studied by Fishbone
\& Moncrief (1976), Abramowicz et al. (1978), Paczy\'nski \& Wiita
(1980), and others (see Frank et al. 1992), and on which the ion torus
model of Rees et al. (1982) is based.

\begin{figure}
\centerline{
\psfig{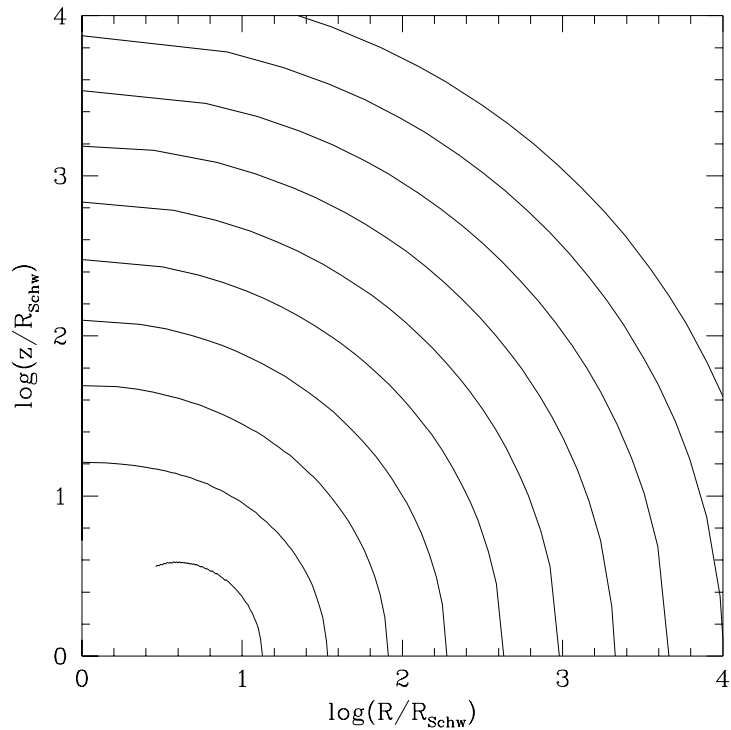}
}
\caption{Isodensity contours (two contours per decade) in the $R-z$
plane for a global ADAF with $\alpha=0.3$, $\gamma=1.4444$ and $f=1$.
The contours are truncated at the sonic radius since the vertical
structure is unreliable inside this radius.  (Taken from Narayan
1997).}
\label{shape}
\end{figure}

When $\alpha \gsim 0.01$, on the other hand, the flow dynamics is
dominated by viscosity, which efficiently removes angular momentum
from the gas.  The gas then has a large radial velocity, $v\sim \alpha
v_{ff}$, remains sub--Keplerian at all radii, and has no pressure
maximum outside the sonic radius.  High $\alpha$ ADAFs therefore do
not possess empty funnels and are unlikely to be toroidal in
morphology.  Instead the flow is probably quasi--spherical all the way
down to the sonic radius; in fact, these ADAFs are more akin to slowly
rotating settling stars, as illustrated in Fig. \ref{shape}.  As we
discuss below (\S 4.1.3), two--temperature ADAFs have $\alpha \sim
0.2-0.3$, which places them firmly in the regime of quasi--spherical,
rather than toroidal, flows.

\subsection{Relativistic Global Solutions}

The global solutions described above are based on Newtonian physics.
Recently, fully relativistic global solutions in the Kerr geometry
have been calculated by Abramowicz et al. (1996), Peitz \& Appl (1997)
and Gammie \& Popham (1998).  Gammie \& Popham (1997) find that their
relativistic solutions are similar to the global solutions of Narayan
et al. (1997a) for radii $R \gsim 10 R_S$, while for $R \lsim 10 R_S$
the solutions differ significantly.  They also show that, in the inner
regions of the flow, the black hole's spin has a substantial effect on
the density, temperature, angular momentum and radial velocity of the
accreting gas.  This is likely to have an impact on the observed
spectrum (Jaroszy\'nski \& Kurpiewski 1997).  A detailed survey of
solutions for various parameters is discussed in Popham \& Gammie
(1998).

To summarize the last four subsections, a fairly advanced level of
understanding has been achieved for the dynamics of ADAFs.  There has
been progress on two fronts.  First, the two-dimensional structure of
the flow in $R-\theta$ is understood in the radially self-similar
regime.  Second, using height-integration (which is shown to be an
excellent approximation in the self-similar zone), the radial
structure of the flow, including the effects of boundary conditions,
has been worked out including full general relativity.  Several
problems, however, remain.  The two-dimensional structure in the
boundary regions, where the flow deviates substantially from
self-similarity, is not understood.  There may be solutions that are
qualitatively distinct from the ones discovered so far, e.g.,
solutions with jets.  In addition, three dimensional effects may arise
from spiral modes, convective turbulence, etc..  Finally, the
introduction of MHD effects in the fluid equations may reveal new
phenomena which go beyond the hydrodynamic approximations considered
so far.

\subsection{On the Possibility of Radial Shocks}

The global steady state solutions discussed in \S 2.4 and \S 2.5 are
free of radial shocks.  Igumenshchev et al. (1997) have carried out
time--dependent simulations of global transonic flows and again find
no shocks.  However, Chakrabarti and his collaborators (see
Chakrabarti 1996 for a review of this work) have claimed that shocks
are generic to ADAFs.  The clearest statement of his results is found
in Chakrabarti \& Titarchuk (1995); the authors claim: (i) Low
$\alpha$ ($\lsim 0.01$) flows have shocks and high $\alpha$ ($\gsim
0.01$) flows do not. (ii) Low $\alpha$ flows have sub--Keplerian
rotation at large radii while high $\alpha$ flows are Keplerian at all
radii except very close to the black hole.  Neither statement has been
confirmed by other workers.  In particular, no shocks are seen in any
of the global solutions, which span a wide range of parameter values:
$\alpha$ ranging from 10$^{-3}$ to 0.3, and $\gamma$ from 4/3 to 5/3.
The reader is referred to Narayan et al. (1997a) and Narayan (1997)
for a more detailed discussion of the lack of radial shocks in ADAF
solutions.

\subsection{ADAFs versus Pure Spherical Accretion}

Geometrically, ADAFs are similar to spherical Bondi accretion.  They
are quasi--spherical and have radial velocities which are close to
free fall, at least for large $\alpha$.  Also, the gas passes through
a sonic radius and falls supersonically into the black hole.

Despite these similarities, it is important to stress that ADAFs are
{\em dynamically} very different from pure spherical accretion.
Transport of angular momentum through viscosity plays a crucial role
in ADAFs; indeed, there would be no accretion at all without viscosity
(recall that $v\propto \alpha$, cf. eq. [\ref{simpleeq}]).  Bondi
accretion, on the other hand, involves only a competition between
inward gravity and outward pressure gradient; viscosity plays no
role.

In the Bondi solution, the location of the sonic radius is determined
by the properties of the accreting gas at large radii.  Under many
conditions, the sonic radius is quite far from the black hole.  In an
ADAF, on the other hand, the sonic radius is almost always at a few
$R_S$.  Furthermore, the location of the sonic radius depends on the
viscosity parameter $\alpha$ (another indication of the importance of
viscosity) and the properties of the gas at infinity are irrelevant so
long as the gas has enough angular momentum to prevent direct radial
infall.  In addition, the gas temperature, or energy density, in an
ADAF is determined by viscous heating and adiabatic compression, and
is generally much higher than in Bondi accretion.  Also, the
rotational velocities in ADAFs are quite large, reaching nearly virial
velocities, whereas $\Omega$ is strictly zero in the Bondi solution.

In nature, when gas accretes onto a black hole, it nearly always has
too much angular momentum to fall directly into the hole.  ADAFs are,
therefore, much more relevant than the Bondi solution for the
description of accretion flows around black holes.
 
\section{Two--Temperature ADAFs}

As mentioned in \S1, two types of ADAFs are known, a high--$\dot{M}$
optically thick solution and a low--$\dot{M}$ optically--thin,
two--temperature solution.  The high--$\dot{M}$ ADAF, though
well--understood dynamically, has seen few applications to
observations.  In particular, the spectral properties of this solution
have not been studied in detail (see, however, Szuszkiewicz et
al. 1996 for a recent study).  In contrast, the low--$\dot{M}$ ADAF
has been extensively studied during the last few years and has been
widely applied to black hole X--ray binaries and low--luminosity
galactic nuclei.  The rest of this article is devoted to this
solution.

The two-temperature ADAF solution (though not by this name) was first
described in a remarkable paper by Ichimaru (1977).  This paper
clearly distinguished the ADAF solution from the SLE solution and
argued (without proof) that the ADAF solution would be stable.  It
also suggested that the black hole X-ray binary Cyg X-1, in its so
called low state (the hardest spectrum in Fig. 14), corresponds to an
ADAF.  Unfortunately, this paper was soon forgotten; in fact, its very
existence was unknown to later generations of disk theorists until the
paper was rediscovered and brought to the attention of the community
by Roland Svensson at the present workshop.

Elements of the two-temperature ADAF were described independently by
Rees et al. (1982) in their ``ion torus'' model.  This paper, however,
though not forgotten, did not have a strong influence on later
workers.  Indeed, all through the 1980's and the early half of the
1990's, the SLE solution was studied in greater detail than the ion
torus model. (Even Svensson did most of his work on the SLE solution!)
It appears that most scientists in the community did not recognize
that the SLE solution and the ion torus model were distinct, and even
those that did were not fully aware of their different stability
properties.

The current interest in the two-temperature ADAF solution was
initiated by the papers of Narayan \& Yi (1994, 1995a, 1995b),
Abramowicz et al. (1995), Chen (1995), Chen et al. (1995), and a host
of others.  These workers discovered the solution for the third time,
but this time developed it in considerable detail, studied its
properties, and applied it widely to a number of systems.

\subsection{Basic Assumptions}

Models based on the two--temperature ADAF solution make certain
critical assumptions.  The validity of these assumptions is not proved
and is currently under investigation.

\subsubsection{Equipartition Magnetic Fields}

It is assumed that magnetic fields contribute a constant fraction
$(1-\beta)$ of the total pressure: \beq p_{\rm m} = {B^2 \over 24 \pi}
= (1-\beta) \rho c_s^2, \eeq where $p_{\rm m}$ is the magnetic
pressure due to an isotropically tangled magnetic field. Note that the
usual $\beta$ of plasma physics is related to the $\beta$ utilized
here by $\beta_{\rm plasma} = \beta/3(1-\beta)$ (the $1/3$ arises
because the plasma $\beta$ uses $B^2/8 \pi$ for the magnetic pressure,
rather than $B^2/24 \pi$).  The assumption of a constant $\beta$ is
fairly innocuous since, in general, we expect equipartition magnetic
fields in most astrophysical plasmas. In particular, Balbus \& Hawley
(1991) have shown that differentially rotating disks with weak
magnetic fields develop a strong linear MHD instability which
exponentially increases the field strength to near equipartition
values.  The exact saturation value of $p_{\rm m}$ is, however,
unclear (e.g., Hawley, Gammie, \& Balbus 1996). ADAF models assume
$\beta=0.5$ (or $\beta_{\rm plasma} = 1/3$), i.e., strict
equipartition between gas and (tangled) magnetic pressure.  The
presence of equipartition magnetic fields implies that the effective
adiabatic index in the energy equation is not that of a monatomic
ideal gas.  Esin (1997) argues that the appropriate expression is
$\gamma = (8 - 3 \beta)/(6 - 3 \beta)$.

\subsubsection{Thermal Coupling Between Ions and Electrons}

ADAF models assume that ions and electrons interact only through
Coulomb collisions and that there is no non--thermal coupling between
the two species.  In this case the plasma is two temperature, with the
ions much hotter than the electrons.  This important assumption may be
questioned on the grounds that magnetized collisionless plasmas, such
as ADAFs, have many modes of interaction; intuitively, it would seem
that the plasma might be able to find a more efficient way than
Coulomb collisions of exchanging energy between the ions and the
electrons (Phinney 1981).  To date, however, only one potential
mechanism has been identified.  Begelman \& Chiueh (1989) show that,
under appropriate conditions, large ion drift velocities (due to
significant levels of small scale turbulence in the accretion flow)
may drive certain plasma waves unstable, which then transfer energy
directly from the ions to the electrons. Narayan \& Yi (1995b),
however, argue that for most situations of interest, the specific
mechanism identified by Begelman \& Chiueh (1989) is not important for
ADAF models.

\subsubsection{Preferential Heating of Ions}

The two-temperature ADAF model assumes that most of the turbulent
viscous energy goes into the ions (SLE; Ichimaru 1977; Rees et
al. 1982; Narayan \& Yi 1995b), and that only a small fraction $\delta
\ll 1$ goes into the electrons. The parameter $\delta$ is generally
set to $\sim 10^{-3} \sim m_e/m_p$, but none of the results depend
critically on the actual value of $\delta$, so long as it is less than
a few percent.  Recently, there have been several theoretical
investigations which consider the question of particle heating in
ADAFs.

Bisnovatyi--Kogan \& Lovelace (1997) argue that large electric fields
parallel to the local magnetic field preferentially accelerate
electrons (by virtue of their smaller mass), leading to $\delta \sim
1$.  In magnetohydrodynamics (MHD), however, an electric field
parallel to the magnetic field arises only from finite resistivity
effects and is of order $\sim vB/c$Re$_{\rm m}$, where Re$_{\rm m} \gg
1$ is the magnetic Reynolds number. Local parallel electric fields are
therefore negligible.  Bisnovatyi--Kogan \& Lovelace (1997) suggest
that, instead of the microscopic resistivity, one must use the
macroscopic or turbulent resistivity to determine the local electric
field, in which case parallel electric fields are large, $\sim vB/c$.
The problem with this analysis, as discussed by Blackman (1998) and
Quataert (1998), is that the turbulent resistivity is defined only for
the large scale (mean) fields.  On small scales, the MHD result
mentioned above is applicable and so parallel electric fields are
unimportant for accelerating particles.

Blackman (1998), Gruzinov (1998), and Quataert (1998) consider
particle heating by MHD turbulence in ADAFs.  Blackman shows that
Fermi acceleration by large scale magnetic fluctuations associated
with MHD turbulence preferentially heats the ions in two temperature,
gas pressure dominated, plasmas.  This is, however, equivalent to
considering the collisionless damping of the fast mode component of
MHD turbulence (Achterberg 1981), and thus does not apply to the
non-compressive, i.e., Alfvenic, component (which is likely to be
energetically as or more important than the compressive component).

Following the work of Goldreich \& Sridhar (1995) on Alfvenic
turbulence, Gruzinov (1998) and Quataert (1998) show that, when the
magnetic field is relatively weak, the Alfvenic component of MHD
turbulence is dissipated on length scales of order the proton Larmor
radius.  The damping is primarily by ``magnetic'' Landau damping (also
known as transit time damping, Cherenkov damping, etc.), not the
cyclotron resonance (which is unimportant), and most of the Alfvenic
energy heats the ions rather than the electrons.  For equipartition
magnetic fields, Gruzinov argues that Alfvenic turbulence will cascade
to length scales much smaller than the proton Larmor radius and heat
the electrons.  An extension of this work is given in Quataert \&
Gruzinov (1998).

\subsubsection{$\alpha$ Viscosity}

The viscosity parameter $\alpha$ of Shakura \& Sunyaev (1973) is used
to described angular momentum transport; $\alpha$ is assumed to be
constant, independent of radius. Some authors have proposed that
$\alpha$ may vary as a function of $(H/R)$.  Since ADAFs have $H\sim
R$, no radial dependence is expected, and a constant $\alpha$ appears
to be a particularly good assumption (Narayan 1996b).  Since the
viscous stress is probably caused by magnetic fields, the parameters
$\alpha$ and $\beta$ are likely to be related according to the
prescription of Hawley, Gammie \& Balbus (1996): $\alpha= 3 c
(1-\beta)/(3 - 2 \beta)$, where $c \sim 0.5-0.6$.  This expression for
$\alpha(\beta)$ differs from the expression given elsewhere in the
literature (e.g., Narayan et al. 1998), $\alpha = c (1 - \beta)$, in
properly recognizing that Hawley et al. define magnetic pressure as
$B^2/8 \pi$, while we use equation (3.14).  For equipartition magnetic
fields ($\beta=0.5$), $\alpha \sim 0.4$.  The models in the literature
usually set $\alpha$ to $0.25$ or $0.3$.

\subsection{Properties of Two--Temperature, Optically Thin, ADAFs}

For the remainder of this review, we write all quantities in scaled
units: the mass is scaled in solar mass units,
$$
M = m M_{\odot},
$$ 
the radius in Schwarzschild radii,
$$
R = r R_{S}, \ \ \ \ \ R_S = {2 GM \over c^2} = 2.95\times 10^5 m  
\ \ \ \mbox{cm},
$$
and the accretion rate in Eddington units,
$$
\dot{M} = \dot{m} \, \dot{M}_{\rm Edd}, \ \ \ 
\dot{M}_{\rm Edd}= {L_{\rm Edd} \over \eta_{\rm eff} c^2} = 1.39\times 
10^{18} \, m \ \ \ \mbox{g s$^{-1}$}, 
$$
where we have set $\eta_{\rm eff}$, the efficiency of converting matter
to radiation, equal to 0.1 in the definition of $\dot{M}_{\rm Edd}$
(cf. Frank et al. 1992).

\subsubsection{Scaling Laws}

Using the self--similar solution (\ref{soln6})--(\ref{soln8}), one can
obtain a fairly good idea of the scalings of various quantities in an ADAF
as a function of the model parameters.  Setting $f\rightarrow 1$
(advection--dominated flow) and $\beta=0.5$ (equipartition magnetic
field), one finds (Narayan \& Yi 1995b; see also Mahadevan 1997),
$$
v \simeq -1.1\times 10^{10}\, \alpha r^{-1/2} \ \ \ \mbox{cm s$^{-1}$,}
$$
$$
\Omega \simeq 2.9 \times 10^{4} \ m^{-1}r^{-3/2} \ \ \ \mbox{s},
$$
$$
c_s^2 \simeq 1.4 \times 10^{20} \ r^{-1} \ \ \ \mbox{cm$^2$ s$^{-2}$},
$$
$$
n_e \simeq 6.3 \times 10^{19}\ \alpha^{-1} \, m^{-1}
	\dot{m} \, r^{-3/2} \ \ \ \mbox{cm$^{-3}$,}
$$
$$
B \simeq 7.8\times 10^{8} \ \alpha^{-1/2} \, m^{-1/2} \dot{m}^{1/2} \,
	r^{-5/4} \ \ \ \mbox{G,}
$$
$$ 
p \simeq 1.7\times 10^{16} \ \alpha^{-1} \, m^{-1}\dot{m} \, r^{-5/2}
 	\ \ \ \mbox{g cm$^{-1}$ s$^{-2}$},
$$
$$
q^+ \simeq 5.0\times 10^{21} \ m^{-2} \dot{m} \, r^{-4} \ \ \ 
	\mbox{ergs cm$^{-3}$ s$^{-1}$},
$$
\beq
\tau_{\rm es} \simeq 24 \ \alpha^{-1} \dot{m} \, r^{-1/2},
\label{scalinglaweqs}
\eeq where $n_e$ is the electron density, $p$ is the pressure (gas
plux magnetic), and $\tau_{\rm es}$ is the electron scattering optical
depth to infinity.

\subsubsection{Critical Mass Accretion Rate}

The optically thin ADAF solution exists only for $\dot{m}$ less than a
critical value $\dot{m}_{\rm crit}$ (Ichimaru 1977; Rees et al. 1982;
Narayan \& Yi 1995b; Abramowicz et al. 1995; Esin et al. 1996, 1997).
To derive this result, consider first an optically thin
one--temperature gas which cools primarily by free--free emission
(Abramowicz et al. 1995).  This is a reasonable model at large radii,
$r>10^{3}$.  The viscous heating rate varies as $q^+ \propto
m^{-2}\dot{m}r^{-4}$ (eq. \ref{scalinglaweqs}), while the cooling
varies as $q^- \propto n_p n_e T_e^{1/2}\propto \alpha^{-2} m^{-2}
\dot{m}^2 T_e^{1/2} r^{-3}$, where $n_p$ is the proton number
density. For a one--temperature ADAF, $T_e \sim m_p c_s^2/k \sim
10^{12}$K$/r$, and is independent of $\dot{m}$. By comparing $q^+$ and
$q^-$, it is easily seen that there is an $\dot{m}_{\rm crit} \propto
\alpha^2 r^{-1/2}$, such that for $\dot{m} < \dot{m}_{\rm crit}$, we
have $q^+ > q^-$ and a consistent ADAF solution, whereas for $\dot{m}
> \dot {m} _{\rm crit}$ no ADAF is possible.  Note that $\dot m_{\rm
crit}$ decreases with increasing $r$.  The decrease is as $r^{-1/2}$
if the cooling is dominated by free--free emission.  If other atomic
cooling processes are also included, the decrease is more rapid
(cf. Fig. \ref{mdotcrit}).

The above argument assumes free--free cooling, which is the dominant
cooling mechanism at non--relativistic temperatures. However, once
electrons become relativistic, other cooling processes such as
synchrotron radiation and inverse Compton scattering take over.  Esin
et al. (1996) showed that even in this situation it is possible to
have a one-temperature ADAF at sufficiently low $\dot{m}$.  However,
the critical accretion rate $\dot{m}_{\rm crit}$ is very small, $\sim
10^{-6}$ (for $\alpha \approx 0.25$; Esin et al. found $\dot{m}_{\rm
crit} \sim 10^{-4} \alpha^2$, but more detailed calculations give a
somewhat smaller value).  Models with such low mass accretion rates
are not of much interest for modeling observed systems.

If we allow a two--temperature plasma, then the ADAF solution can,
utilizing the assumptions given in the previous subsections, extend up
to larger and more interesting mass accretion rates (Ichimaru 1977;
Rees et al. 1982; Narayan \& Yi 1995b).  By assumption, the viscous
energy primarily heats the protons, while the cooling is almost
entirely by the electrons.  At low densities Coulomb coupling between
protons and electrons is very weak and the amount of viscous energy
that is transferred to the electrons is very small.  Coulomb coupling
therefore acts as a bottleneck which restricts the amount of energy
that can be lost to radiation.

With increasing $\dot{m}$, Coulomb coupling becomes more efficient,
and at a critical density the coupling is so efficient that a large
fraction of the viscous energy is transferred to the electrons and is
radiated.  Above this accretion rate, the flow ceases to be an ADAF;
it becomes a standard cooling--dominated thin disk.  The critical
accretion rate can be estimated by determining the $\dot{m}$ at which
the viscous heating, $q^+$, equals the rate of energy transfer from
the ions to the electrons, $q^{\rm ie}$.  Alternatively, we can set
the ion--electron equilibration time (the time for collisions to force
$T_i \approx T_e$), $t_{\rm ie}$, equal to the accretion time, $t_{\rm
a}$.  The former time scale is given by (Spitzer 1962)
\begin{eqnarray}
t_{\rm ie} &=& {(2\pi)^{1/2} \over 2 \, n_e \sigma_T c \ln \Lambda} \,
\left( {m_p \over m_e} \right) \, (\theta_e + \theta_p)^{3/2},
\nonumber \\ &\simeq& 9.3 \times 10^{-5} {\alpha } \, \theta_e^{3/2}
\, m \, \dot{m}^{-1} \, r^{3/2} \ \ \ {\rm s}, \nonumber
\end{eqnarray}
where $\ln \Lambda \sim 20$ is the Coulomb logarithm, $\theta_p=
kT_p/m_pc^2$, and $\theta_e = kT_e/m_ec^2$.  The accretion time is
given by \beq t_{\rm a} = \int {dR \over v(R)} \simeq 1.8\times
10^{-5} \, {\alpha}^{-1} \, m \, r^{3/2} \ \ \ \mbox{ s}. \nonumber
\eeq Setting these two timescales equal gives \beq \dot{m}_{\rm crit}
\simeq 5 \theta_e^{3/2} \alpha^2 \simeq 0.3 \alpha^2, \eeq where we
have used $\theta_e \sim 0.16$, corresponding to $T_e = 10^9$K
(Mahadevan 1997).  More detailed models (cf. Esin et al. 1997) give
$\dot{m}_{\rm crit} \sim \alpha^2$.  This value of $\dot m_{\rm crit}$
is essentially independent of $r$ out to about $10^2-10^3$
Schwarzschild radii.  Beyond that, the accreting gas becomes
one-temperature and $\dot m_{\rm crit}$ decreases with increasing $r$,
as explained above.  Figure (\ref{mdotcrit}) shows a detailed estimate
of the profile of $\dot m_{\rm crit}$ vs $r$.  In this review, we will
refer to the full profile as $\dot m_{\rm crit}(r)$ and refer to $\dot
m_{\rm crit}(r_{ms})$ as simply $\dot m_{\rm crit}$, where $r_{ms}$ is
the marginally stable orbit ($r_{ms}=3$).  Thus, $\dot m_{\rm crit}$
refers to the maximum $\dot m$ up to which an ADAF zone of any size is
allowed.

As we discuss later, observations suggest that the two--temperature
ADAF solution exists up to $\dot{m}_{\rm crit}\sim 0.05-0.1$.  This
suggests that $\alpha \sim 0.2-0.3$ in ADAFs.

\subsubsection{Temperature Profiles of Ions and Electrons}

In a two--temperature ADAF the ions receive most of the viscous energy
and are nearly virial (Narayan \& Yi 1995b),
$$
T_i \simeq 2 \times 10^{12}  \beta r^{-1}.
$$
The electrons, on the other hand, are heated by several processes with
varying efficiencies --- Coulomb coupling with the ions, 
compression, and direct viscous heating (described by the parameter
$\delta$) --- and cooled by a variety of radiation processes.

To determine the electron temperature profile, consider the electron
energy equation (Nakamura et al. 1996, 1997; Mahadevan \& Quataert
1997),
\beq \rho T_e v {ds \over dR} = \rho v {d\epsilon \over
dR} - q^c = q^{ie} + q^v - q^-,
\qquad q^c \equiv kTv {dn \over dR}, 
\label{electronenergyeq}
\eeq where $s$ and $\epsilon$ are the entropy and internal energy of
the electrons per unit mass of the gas, and $q^c$ and $q^-$ are the
compressive heating (or cooling) rate and the energy loss due to
radiative cooling per unit volume.  The total heating rate of the
electrons is the sum of the heating via Coulomb collisions with the
hotter protons, $q^{\rm ie}$, and direct viscous heating, $q^v =
\delta q^+$.  (Recall that $\delta$ is usually assumed to be
$\sim10^{-3}$ in ADAF models.)

\begin{figure}
\centerline{
\psfig{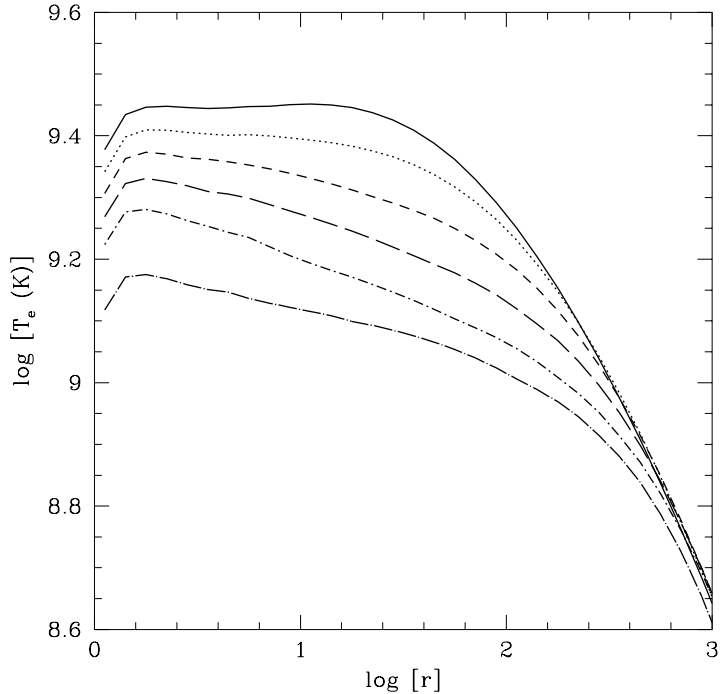}
}
\caption{Variation of electron temperature $T_e$ with radius $r$ for
ADAFs with (from top to bottom) $\log(\dot m) = -2, -1.8, -1.6, -1.4,
-1.2, -1.1$ (taken from Esin et al. 1997).  Note that the electron
temperature decreases with increasing $\dot m$.}
\label{tefromesin}
\end{figure}

The electron temperature at radii $r\lsim 10^2$ in two--temperature
ADAFs is generally in the range $10^9-10^{10}$K. The exact form of the
electron temperature profile depends on which of the three heating
terms dominates.  Nakamura et al. (1996, 1997) were the first to
emphasize the compressive heating term, $q^c$.  Mahadevan \& Quataert
(1997) showed that compressive heating of the electrons is more
important than direct viscous heating so long as $\delta \lsim
10^{-2}$.  Since $\delta \sim 10^{-3}$ in most models, we can ignore
direct viscous heating and consider only the following two limits.
\begin{enumerate}
\item[$q^{\rm ie}\gg q^c$:] This condition is satisfied for 
$\dot{m} \gsim 0.1 \alpha^2$.  Rewriting equation
(\ref{electronenergyeq}) in this regime gives 
$$
\rho v{d\epsilon \over dR} \simeq q^{\rm ie} + q^{\rm c} - q^-
\simeq q^{\rm ie} - q^-.
$$ 
Since the cooling is efficient enough to radiate away all of the
energy given to the electrons, $q^{\rm ie} = q^-$ for $r \lsim 10^2$,
and the internal energy of the electrons does not change with radius,
$d\epsilon/dR = 0$.  The electron temperature therefore remains
essentially constant at $\sim10^9$K (e.g. Narayan \& Yi 1995b).  The
value of $T_e$ depends weakly on $\dot m$ (Fig. \ref{tefromesin}).
\item[$q^{c} \gg q^{\rm ie}$:] This condition is satisfied for
$\dot{m} \lsim 10^{-4} \alpha^2$. Rewriting equation
(\ref{electronenergyeq}) in this regime gives
$$ 
\rho v{d\epsilon \over dR} \simeq q^{\rm c} + q^{\rm ie} - q^-
\simeq q^{\rm c}.
$$ 
The radial dependence of the electron temperature is determined by
adiabatic compression (Nakamura et al. 1997; Mahadevan \& Quataert
1997).  The electron temperatures in this regime are slightly higher.
\end{enumerate}

For $10^{-4}\alpha^2 \leq \dot{m} \leq 0.1 \alpha^2$, the electron
temperature profile lies in between these two extremes, and is
determined by solving equation (\ref{electronenergyeq}) without
neglecting any of the heating terms.

Figure \ref{tefromesin} shows the temperature profile as a function of
radius for various $\dot{m}$.  The electrons actually become cooler as
the accretion rate increases.  Since with increasing $\dot{m}$ the
energy transferred to the electrons goes up, one might naively expect
the electrons to reach higher temperatures in order to radiate the
additional energy.  In fact, the opposite happens; at high $\dot m$,
the dominant cooling mechanism is inverse Compton scattering.  This is
an extremely efficient process when the optical depth approaches unity
and its efficiency increases sharply with increasing optical depth,
i.e., increasing $\dot m$.

Although the electrons achieve relativistic temperatures, pair
processes are found to be unimportant in two-temperature ADAFs
(Kusunose \& Mineshige 1996, Bj\"ornsson et al. 1996, Esin et al. 1997).
The reason is the low density, which allows very few pair-producing
interactions in the medium.

\subsubsection{ADAF Spectra and Radiation Processes}

\begin{figure}
\centerline{
\psfig{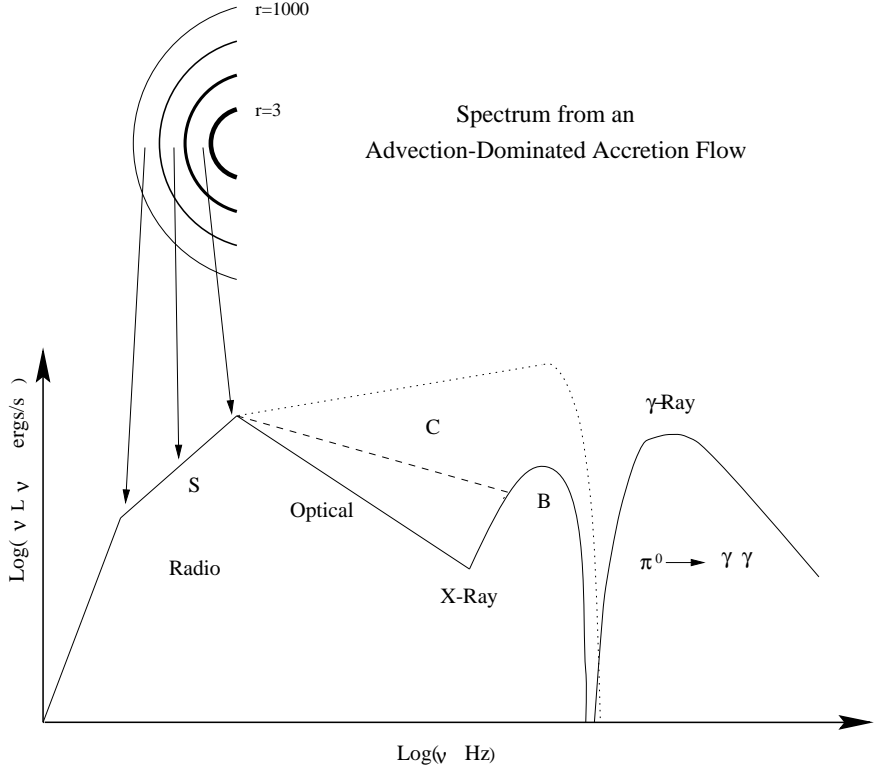}
}
\caption{Schematic spectrum of an ADAF around a black hole.  S, C, and
B refer to electron emission by synchrotron radiation, inverse Compton
scattering, and bremsstrahlung, respectively.  The solid line
corresponds to a low $\dot m$, the dashed line to an intermediate
$\dot m$, and the dotted line to a high $\dot m \sim \dot m_{\rm
crit}$.  The $\gamma$--ray spectrum is due to the decay of neutral
pions created in proton-proton collisions.}
\label{schematic}
\end{figure}

The spectrum from an ADAF around a black hole ranges from radio
frequencies $\sim 10^9$Hz to gamma--ray frequencies $\gsim 10^{23}$Hz,
and can be divided into two parts based on the emitting particles: (1)
The radio to hard X--ray radiation is produced by electrons via
synchrotron, bremsstrahlung and inverse Compton processes (Mahadevan
1997).  (2) The gamma--ray radiation results from the decay of neutral
pions created in proton--proton collisions (Mahadevan, Narayan \&
Krolik 1997).  Figure (\ref{schematic}) shows schematically the
various elements in the spectrum of an ADAF around a black hole.

\begin{figure}
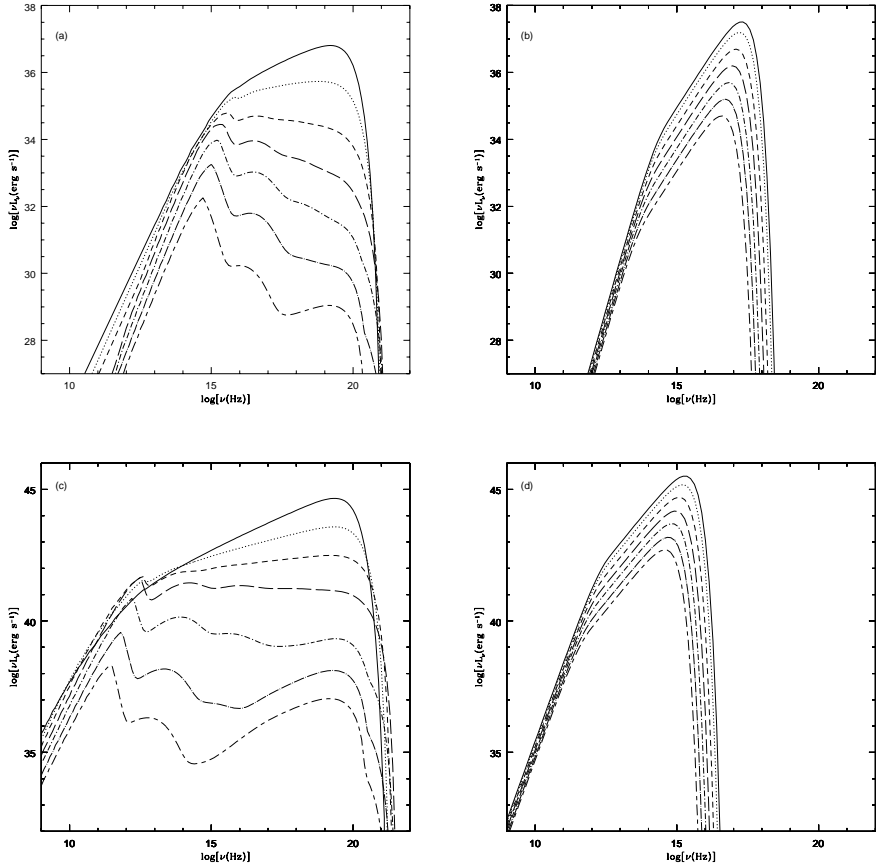

\begin{center}
\begin{minipage}[t]{0.45\hsize}
\begin{displaymath}
\psfig{figure=nfig6a.epsi,height=2.3in}
\end{displaymath}
\end{minipage}
\begin{minipage}[t]{0.45\hsize}
\begin{displaymath}
\psfig{figure=nfig6b.epsi,height=2.3in}
\end{displaymath}
\end{minipage}
\begin{minipage}[t]{0.45\hsize}
\begin{displaymath}
\psfig{figure=nfig6c.epsi,height=2.3in}
\end{displaymath}
\end{minipage}
\begin{minipage}[t]{0.45\hsize}
\begin{displaymath}
\psfig{figure=nfig6d.epsi,height=2.3in}
\end{displaymath}
\end{minipage}
\end{center}
\caption{(a) Spectra from an ADAF around a 10 solar mass black hole
for (from top to bottom) log$(\dot m) = \log(\dot m_{\rm crit})
\approx -1.1, -1.5, -2, -2.5, -3, -3.5, -4$.  (b) Spectra from a thin
disk at the same accretion rates.  Figures (c) and (d) show the
corresponding spectra for a $10^9$ solar mass black hole.  Note that
these spectra are for pure disk and pure ADAF models.  In practice,
real systems are often modeled as an ADAF surrounded by a thin disk
(see \S 3.3).  In such composite models, the ADAF part of the spectrum
is essentially unchanged, but a dimmer and softer version of the thin
disk is also present in the spectrum.}
\label{spectra}
\end{figure}

The low energy end of the spectrum, labeled S in the figure, is due to
synchrotron cooling by semi--relativistic thermal electrons (Mahadevan
et al. 1996).  The synchrotron emission is highly self--absorbed and
is very sensitive to the electron temperature ($\nu L_{\nu}\propto
T_e^7$; Mahadevan 1997). The emission at the highest (peak) frequency
comes from near the black hole, while that at lower frequencies comes
from further out.  The peak frequency varies with the mass of the
black hole and the accretion rate, roughly as $\nu^S_{\rm peak}
\propto m^{-1/2} \dot m^{1/2}$ (Mahadevan 1997; see figure
\ref{spectra}).

The soft synchrotron photons inverse Compton scatter off the hot
electrons in the ADAF and produce harder radiation extending up to
about the electron temperature $\sim$ 100 keV ($h \nu^C_{\rm max}
\approx k T_e$).  The relative importance of this process depends on
the mass accretion rate.  At high $\dot m$, when the Compton
$y$-parameter is large (because of the increased optical depth), the
inverse Compton component dominates the spectrum, as shown by the
dotted line labeled C in figure (\ref{schematic}).

As $\dot{m}$ decreases, Comptonization becomes less efficient and the
inverse Compton component of the spectrum becomes softer and less
important (dashed and solid lines labeled C in figure
\ref{schematic}).  At low $\dot{m}$, the X--ray spectrum is dominated
by bremsstrahlung emission, which again cuts off at the electron
temperature ($h\nu^B_{\rm max} \approx kT_e$, the curve labeled B in
figure \ref{schematic}).

Mahadevan et al. (1997) have studied gamma-ray emission from an ADAF
via the decay of neutral pions produced in proton-proton collisions.
The results depend sensitively on the energy spectrum of the protons.
If the protons have a thermal distribution, the gamma-ray spectrum is
sharply peaked at $\sim 70$ MeV, half the rest mass of the pion, and
the luminosity is not very high.  If the protons have a power--law
distribution, the gamma-ray spectrum is a power--law extending to very
high energies (see figure \ref{schematic}), and the luminosity is much
higher; the photon index of the spectrum is equal to the power--law
index of the proton distribution function.

Figure (\ref{spectra}) shows four sequences of spectra, two
corresponding to ADAFs (excluding the gamma-ray component) and two to
thin disks.  Figures (\ref{spectra}a) and (\ref{spectra}b) are for a
$10$ solar mass black hole while Figures (\ref{spectra}c) and
(\ref{spectra}d) are for a $10^9$ solar mass black hole. All sequences
extend over the same range of $\dot m=10^{-4} - 10^{-1.1} \approx \dot
m_{\rm crit}$.  Note how very different the thin disk and ADAF spectra
are, suggesting that it should be easy to tell from spectral
observations whether a system has an ADAF or a thin disk.

Figure (\ref{spectra}) shows that for high $\dot m$, the Compton
component of an ADAF spectrum is roughly a power law while for lower
$\dot m$ distinct Compton peaks are present.  This is primarily due to
the increase of the electron temperature with decreasing $\dot{m}$
(see fig. \ref{tefromesin}).  A larger $T_e$ yields a larger energy
boost per Compton scattering..  At low $\dot m$, the energy gain due
to Compton scattering exceeds the width (in energy) of the input
synchrotron photons, resulting in distinct Compton peaks.  For
increasing $\dot m$, however, $T_e$, and thus the mean energy gain due
to Compton scattering, decreases.  The Compton peaks therefore
``blur'' together, yielding an effective power law.

\begin{figure}
\centerline{
\psfig{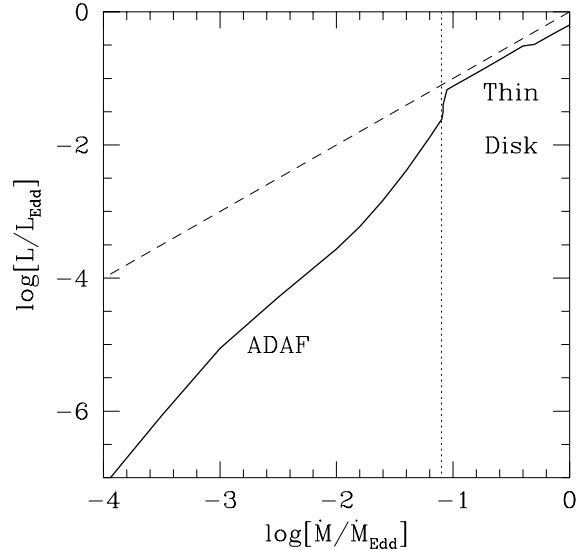}
}
\caption{The bolometric luminosity vs. mass accretion rate according
to the model developed by Esin et al. (1997).  The vertical dotted
line corresponds to $\dot m_{\rm crit}$ (for $\alpha = 0.3$).  Above
this $\dot m$, the accretion is via a thin disk and $L \propto \dot
M$.  Below $\dot m_{\rm crit}$, the accretion is via an ADAF at small
radii and a thin disk at large radii (cf. \S 3.3).  Here $L \propto
\dot M^2$ because much of the viscously generated energy is advected
into the black hole.  The dashed line corresponds to $L = 0.1 \dot M
c^2$.}
\label{lum}
\end{figure}

Another striking feature seen in Fig. (\ref{spectra}) is that ADAFs
are much less luminous than thin disks at low values of $\dot m$.
This is because most of the energy in an ADAF is advected, rather than
radiated, leading to a low radiative efficiency.  In fact, the
luminosity of an ADAF scales roughly as $\sim \dot m^2$ (i.e., the
radiative efficiency scales as $\dot m$).  A thin disk, on the other
hand, has a constant efficiency $\sim10\%$ and the luminosity scales
as $\dot m$.  This is illustrated in Fig. (\ref{lum}) which shows
results of a detailed model described by Esin et al. (1997).

\subsubsection{The Particle Distribution Function in ADAFs}

In determining the radiation processes in ADAFs, the electrons are
assumed to be thermal, while the protons could be thermal or
non--thermal. It is clear that the spectrum will depend significantly
on the energy distribution of the particles.

Mahadevan \& Quataert (1997) considered two possible thermalization
processes in ADAFs: (1) Coulomb collisions and (2) synchrotron
self--absorption.  Using an analytic Fokker--Planck treatment, they
compared the time scales for these processes with the accretion time,
and determined the accretion rates at which thermalization is
possible.  In the case of the protons they found that, for all
accretion rates of interest, neither Coulomb collisions nor
synchrotron self-absorption leads to any significant thermalization.
The proton distribution function is therefore determined principally
by the characteristics of the viscous heating mechanism, and could
therefore be thermal or non--thermal.  Quataert (1998) has argued that
Alfvenic turbulence does {\em not} lead to strong non-thermal features
in the proton distribution function; fast mode turbulence (Fermi
acceleration), however, may generate a power-law tail.

The electrons exchange energy quite efficiently by Coulomb collisions
for $\dot{m} \gsim 10^{-2} \alpha^2$, and are therefore thermal at
these accretion rates. At lower $\dot{m}$ the emission and absorption
of synchrotron photons allows the electrons to communicate with one
another, and therefore to thermalize, even though the plasma is
effectively collisionless (Ghisellini \& Svensson 1990, 1991). At a
radius $r$ in the accretion flow, this leads to thermalization for
$\dot{m} \gsim 10^{-5} \alpha^2 r$ (Mahadevan \& Quataert 1997).  For
still lower $\dot{m}$, the electron distribution function is somewhat
indeterminate, and is strongly influenced by adiabatic compression
(Mahadevan \& Quataert 1997).  Detailed spectra of ADAFs at such low
$\dot{m}$, which take into account the nonthermal distribution
function of the electrons, have not been calculated (see, however,
Fujita et al. (1998) for preliminary models of isolated black holes
accreting at very low $\dot m$ from the interstellar medium).

\subsubsection{Stability of the ADAF Solution}

Thin accretion disks suffer from thermal and viscous instabilities
under certain conditions (e.g., Pringle 1981, Frank et al. 1992).
What are the stability properties of ADAFs?  

Narayan \& Yi (1995b) and Abramowicz et al. (1995) have shown that
ADAFs are stable to long wavelength perturbations (see Narayan 1997
for a qualitative discussion of the relevant physics), while Kato et
al. (1996, 1997) and Wu \& Li (1996) showed that a one temperature
ADAF may be marginally unstable to small scale perturbations.  Using a
time--dependent analysis, Manmoto et al. (1996) confirmed that density
perturbations on small scales in a one-temperature ADAF do grow as the
gas flows in, but not sufficiently quickly to affect the global
validity of the solutions.  They suggest that such perturbations may
account for the variable hard X--ray emission which is observed in AGN
and stellar mass black hole candidates.  Recently, Wu (1997) has
considered the stability of two temperature ADAFs to small scale
fluctuations, and has shown that these flows are both viscously and
thermally stable under most reasonable conditions.

\subsection{ADAF Plus Thin Disk Geometry: The Transition Radius} 

Section 4 describes several applications of the ADAF model to X-ray
binaries and AGN.  Many of these applications utilize the following
geometry, proposed by Narayan, McClintock, \& Yi (1996).  In this
model, the accretion flow consists of two zones separated at a
transition radius, $r_{tr}$.  For $r < r_{tr}$, there is a
two-temperature ADAF, whose properties are described in the previous
two subsections.  For $r > r_{tr}$, the accretion occurs partially as
a thin accretion disk, and partially as a hot corona; the corona is
modeled as an ADAF (Narayan, Barret, \& McClintock 1997; Esin,
McClintock, \& Narayan 1997).  The geometry is very similar to that
proposed by SLE (see also Wandel \& Liang 1991); the main difference
is that the hot phase is taken to be a two-temperature ADAF, rather
than the SLE solution.

Several proposals have been made to explain why the inflowing gas
might switch from a thin disk to an ADAF at the transition radius.
Meyer \& Meyer--Hofmeister (1994) proposed, for cataclysmic variables,
a mechanism in which the disk is heated by electron conduction from a
hot corona; this evaporates the disk, leading to a quasi-spherical hot
accretion flow.  Honma (1996) suggested that turbulent diffusive heat
transport from the inner regions of the ADAF produces a stable hot
accretion flow out to large radii, which then joins to a cool thin
disk.  Narayan \& Yi (1995b) suggested that small thermal
instabilities in the optically-thin upper layers of a thin disk might
cause the disk to switch to an ADAF.  Other ideas are discussed by
Ichimaru (1977) and Igumenshchev, Abramowicz, \& Novikov (1997).  It
is not clear which, if any, of these mechanisms is most important.

\begin{figure}
\centerline{
\psfig{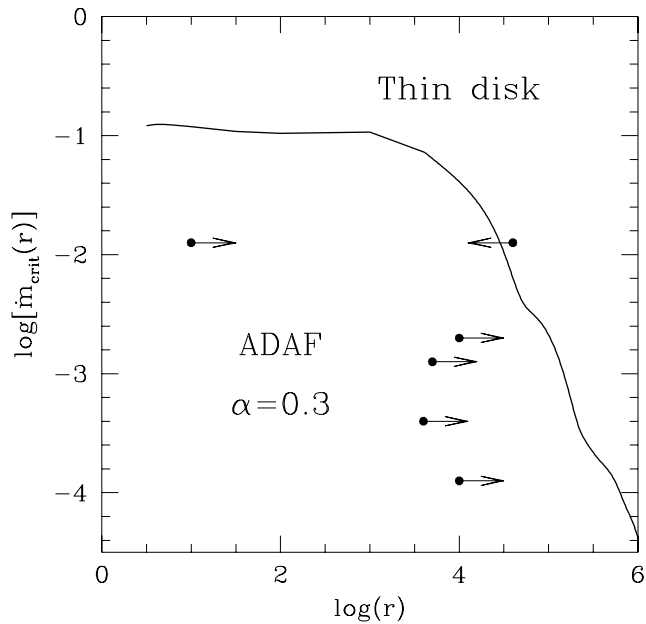}
}
\caption{The solid line shows the estimated variation of $\dot m_{\rm
crit}(r)$ with $r$.  The horizontal segment up to $r=10^3$ is from Esin
et al. (1997) and the curve beyond that is based on a detailed model
of optically thin cooling in a one-temperature gas (Menou et
al. 1998).  According to the ``strong ADAF proposal'' of Narayan \& Yi
(1995b), the curve also shows the variation of $r_{tr}$ with $\dot m$.
Regions above and to the right of the curve correspond to the thin
disk solution, while regions below and to the left correspond to the
ADAF.  The dots and arrows show observationally derived limits on
$r_{tr}$.  The lower limits on $r_{tr}$ are derived from fitting the
spectra of various systems (from above: NGC 4258, V404 Cyg, GRO
J1655-40, A0620--00, and Sgr A$^*$), and also requiring the thin disk
to be thermally stable in the thermal limit cycle model.  The upper
limit on $r_{tr}$ in NGC4258 corresponds to the lowest radius at which
maser emission has been detected.  The calculated curve is consistent
with the various constraints shown, but it is clear that fairly large
changes are allowed, especially for $\dot m>10^{-2.5}$.}
\label{mdotcrit}
\end{figure}

It is generally believed that the transition radius $r_{tr}$ is
determined principally by $\dot m$, but the exact form of $r_{tr}(\dot
m)$ is not known.  Narayan \& Yi (1995b) suggested that, whenever the
accreting gas has a choice between a thin disk and an ADAF, the ADAF
configuration is chosen.  According to this (rather extreme) principle
(the ``strong ADAF principle''), $r_{tr}(\dot m)$ is the maximum $r$
out to which an ADAF is allowed for the given $\dot m$; equivalently,
it is that $r$ at which $\dot m_{\rm crit}(r)=\dot m$ (see
Fig. \ref{mdotcrit}).  This prescription suggests that at low $\dot m
\ll \dot m_{\rm crit}\sim \alpha^2$, $r_{tr}$ will be very large even
though a thin disk is perfectly viable at all radii.  The evidence to
date from quiescent X-ray binaries and galactic nuclei is consistent
with this prediction (\S4 and Fig. \ref{mdotcrit}).

Even if the ``strong ADAF proposal'' of Narayan \& Yi (1995b) is
correct, it does not allow a reliable determination of $r_{tr}(\dot
m)$ because the precise form of $\dot m_{\rm crit}(r)$ is not known.
The plot shown in Fig. (\ref{mdotcrit}) is based on fairly detailed
computations by Esin et al. (1997) and Menou et al. (1998), but it
still makes use of simplifying assumptions that may significantly
influence the results.

\section{Applications}

ADAF models have been applied to a number of accreting black hole
systems.  They give a satisfying description of the spectral
characteristics of several quiescent black hole binaries (Narayan et
al. 1996, 1997b; Hameury et al. 1997) and low luminosity galactic
nuclei (Narayan et al. 1995, 1998; Manmoto et al. 1997; Lasota et
al. 1996; Reynolds et al. 1996; DiMatteo \& Fabian 1997a) which are
known to experience low efficiency accretion.  ADAF models have also
been applied successfully to more luminous systems which have higher
radiative efficiencies (Esin et al. 1997, 1998).

The basic ADAF model has one adjustable parameter, $\dot m$; in
principle, the transition radius $r_{tr}$ is a second parameter, but
most often the results are very insensitive to the choice of $r_{tr}$.
In the future, the Kerr rotation parameter of the black hole will be
an additional parameter of the models, but current models assume a
Schwarzschild black hole.  The mass of the black hole and inclination
of the equatorial plane to the line of sight are estimated from
observations, while the parameters describing the microphysics of the
accretion flow are set to their canonical values (\S3.1):
$\alpha=0.25$ or 0.30, $\beta=0.5$, $\delta=0.001$.  (Any value of
$\delta$ between 0 and 0.01 gives virtually identical results; see
\S3.2.3).

The X-ray flux is very sensitive to the density of the plasma and
therefore to the accretion rate.  For this reason, $\dot m$ is usually
adjusted to fit the observed X-ray flux.  The models described below
are generally in good agreement with the remaining data that are not
used in the fit:  namely, the X-ray spectral slope, the optical/UV
data, and, where available, the radio and infrared observations.  In
contrast, the thin disk model fits the observations very poorly.

The models presented here are the most up to date available, often
more advanced than those in the literature; in particular, they
include the compressive heating of electrons in the electron energy
equation (see \S3.2.3) and the full relativistic dynamics for
accretion onto a Schwarzschild black hole (the photon transport
includes gravitational redshift, but is otherwise Newtonian).

\subsection{Applications to X--Ray Binaries}

\subsubsection{Quiescent Black Hole Transients}

ADAF models have been used to explain the spectra of a number of
quiescent black hole soft X--ray transients (SXTs), namely, A0620--00,
V404 Cyg, and GRO J1655-40 (Narayan et al. 1996; Narayan et al. 1997b;
Hameury et al. 1997).  SXTs are mass transfer binaries in which the
accreting star is often a black hole candidate (though sometimes a
neutron star), and the companion star is usually a low--mass main
sequence star (van Paradijs \& McClintock 1995).  Episodically, these
systems enter a high luminosity, ``outburst,'' phase, but for most of
the time they remain in a very low luminosity, ``quiescent,'' phase.
One problem with modeling quiescent SXTs is that a thin accretion disk
cannot account for both the observed optical/UV and X--ray flux
self--consistently.  For example, in the case of A0620--00, a standard
thin disk requires an accretion rate of
$\dot{M}\sim10^{-10}M_{\odot}$yr$^{-1}$ to explain the optical/UV
flux, while the X--ray flux requires an accretion rate of $\dot{M}
\sim 10^{-15}M_{\odot}$yr$^{-1}$ (McClintock et al. 1995).  Another
problem is that the optical spectrum resembles a blackbody with a
temperature of $\sim10^4$K, but a thin disk cannot exist at such a
temperature since it would be thermally unstable (Wheeler 1996; Lasota
et al. 1996a).

\begin{figure}
\centerline{
\psfig{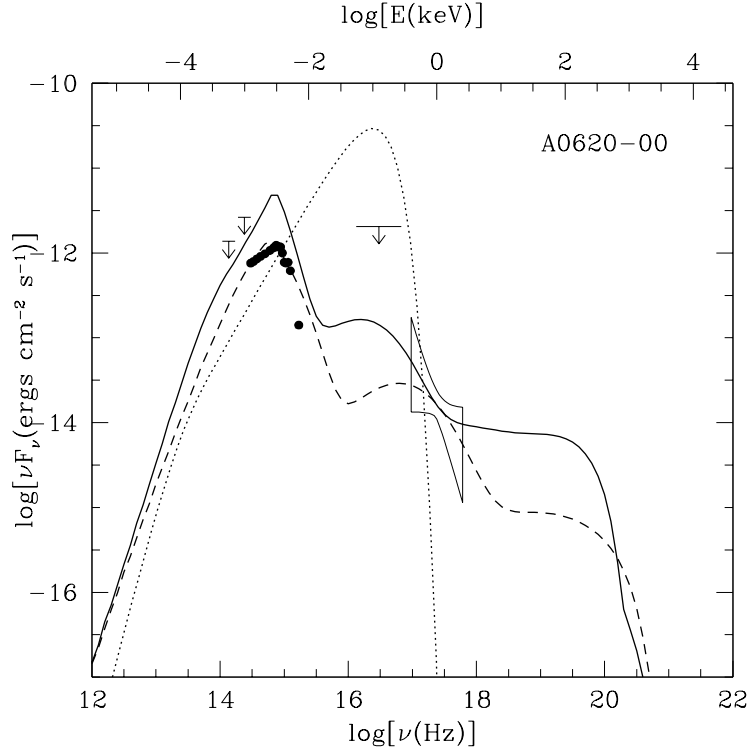}
}
\caption{Spectrum of an ADAF model of A0620-00 (solid line) at an
accretion rate of $\dot m = 4 \times 10^{-4}$, compared with the
observational data.  The dashed line is an ADAF model with $\beta =
0.8$, instead of the standard value of $\beta = 0.5$.  The dotted line
shows the spectrum of a thin accretion disk with an accretion rate
$\dot{m} = 1 \times 10^{-5}$, adjusted to fit the optical flux.
(Based on Narayan et al. 1997b)}
\label{a0620}
\end{figure}

\begin{figure}
\centerline{
\psfig{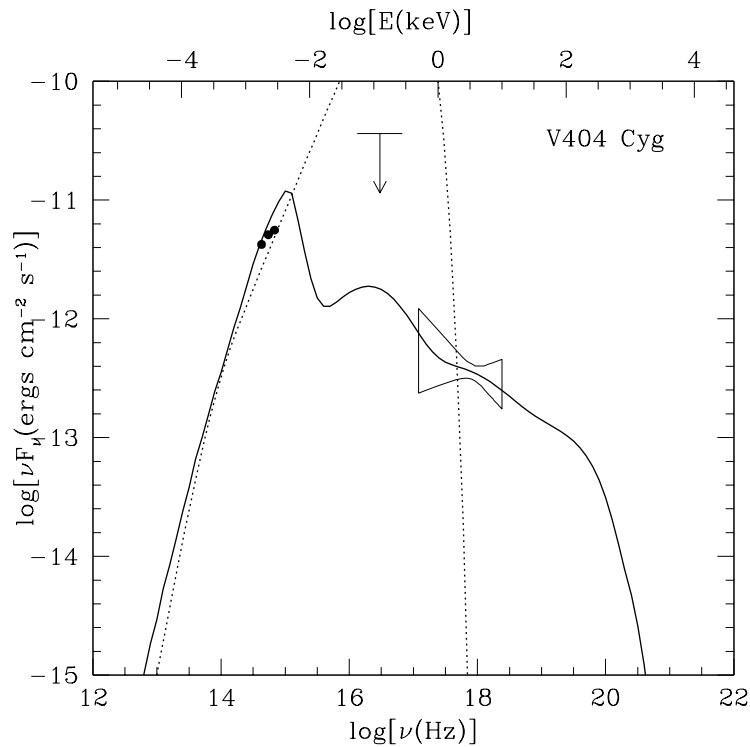}
}
\caption{Spectrum of an ADAF model of V404 Cyg (solid line) at an
accretion rate of $\dot m = 2 \times 10^{-3}$, compared with the
observational data.  The dotted line shows the spectrum of a thin
accretion disk with $\dot{m} = 1.8 \times 10^{-3}$ (adjusted to fit
the optical flux).}
\label{v404}
\end{figure}

Narayan, McClintock, and Yi (1996) and Narayan et al. (1997b) resolved
these problems using an ADAF + thin disk model with $r_{\rm tr} \sim
10^3-10^4$.  The resulting spectra for A0620--00 and V404 Cyg are
shown by solid lines in Figures \ref{a0620} and \ref{v404}.  The ADAF
models of the two sources reproduce the observed X-ray spectral slopes
well; they also reproduce the optical/UV fluxes and spectral shapes
reasonably well (note especially the good agreement in the position of
the optical peak for A0620--00), though the optical flux is generally
somewhat too luminous.  Hameury et al. (1997) showed that observations
of another SXT, GRO J1655-40, are also consistent with the presence of
an ADAF in quiescence.  The dotted lines in Figures (\ref{a0620}) and
(\ref{v404}) are steady state thin disk models adjusted to fit the
optical flux; these models are clearly ruled out by the data.

The optical emission in the ADAF models of SXTs is due to synchrotron
emission from the ADAF.  The thin disk itself is very cool and is in a
stable ``low state.''  There is thus no difficulty with the thermal
instability which a pure thin disk model would face in modeling the
optical spectrum (Wheeler 1996; Lasota et al. 1996a).  The ADAF model
does overproduce the optical flux, but this can be fixed by changing
the value of $\beta$ from 0.5 to 0.8 (see the dashed line in figure
\ref{a0620}).

\subsubsection{Clues From Outburst Timescales}

The above spectral models of quiescent SXTs require an inner ADAF
which connects to an outer thin disk at a large transition radius.
Lasota et al. (1996a) showed that a large transition radius can also
be inferred from the outburst timescales of SXTs.  In addition,
optical and X--ray observations of the black hole SXT GRO J1655--40
showed an outburst in April of 1996.  The optical outburst preceded
the X-ray outburst by roughly 6 days (Orosz et al. 1997), which can be
understood using an ADAF + thin disk model (but not using a pure thin
disk extending down to the last stable orbit).  See Lasota (this
volume) for a detailed discussion of the implications of SXT outbursts
for ADAF models.

\subsubsection{Spectral Models of Luminous X--Ray Binaries}

\begin{figure}
\centerline{ 
\psfig{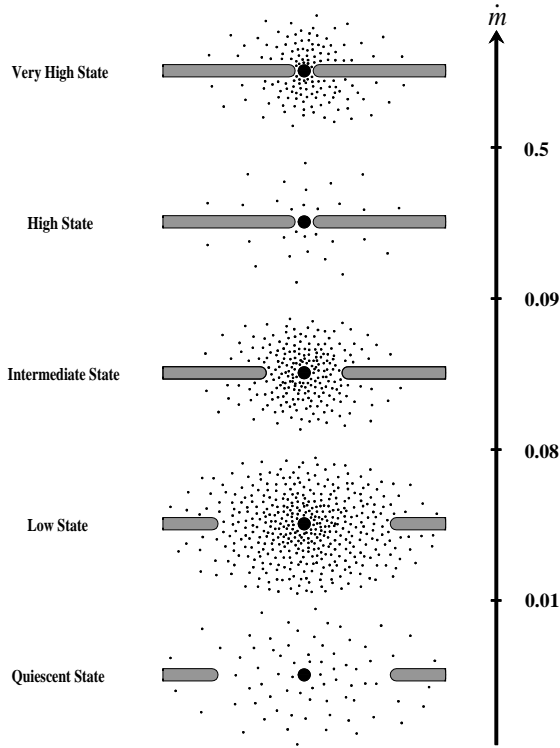} 
}
\caption{The configuration of the accretion flow in different spectral
states shown schematically as a function of the total mass accretion
rate $\dot m$ (from Esin et al. 1997).  The ADAF is indicated by dots
and the thin disk by the horizontal bars.  The lowest horizontal panel
shows the quiescent state which corresponds to a low mass accretion
rate (and therefore, a low ADAF density) and a large transition
radius.  The next panel shows the low state, where the mass accretion
rate is larger than in the quiescent state, but still below the
critical value ${\dot m}_{\rm crit}$.  In the intermediate state (the
middle panel), $\dot m \sim {\dot m}_{\rm crit}$ and the transition
radius is smaller than in the quiescent/low state.  In the high state,
the thin disk extends down to the last stable orbit and the ADAF is
confined to a low-density corona above the thin disk.  Finally, in the
very high state, it has been suggested that the corona may have a
substantially larger $\dot m$ than in the high state, but this is very
uncertain.}
\label{5states}
\end{figure}

Narayan (1996a) proposed that the many different spectral states
observed in black hole X-ray binaries can be understood as a sequence
of thin disk + ADAF models with varying $\dot m$ and $r_{tr}$.  These
ideas have been worked out in more detail by Esin et al. (1997;
1998). A schematic diagram of their model is shown in Figure
(\ref{5states}).

\begin{figure}
\centerline{
\psfig{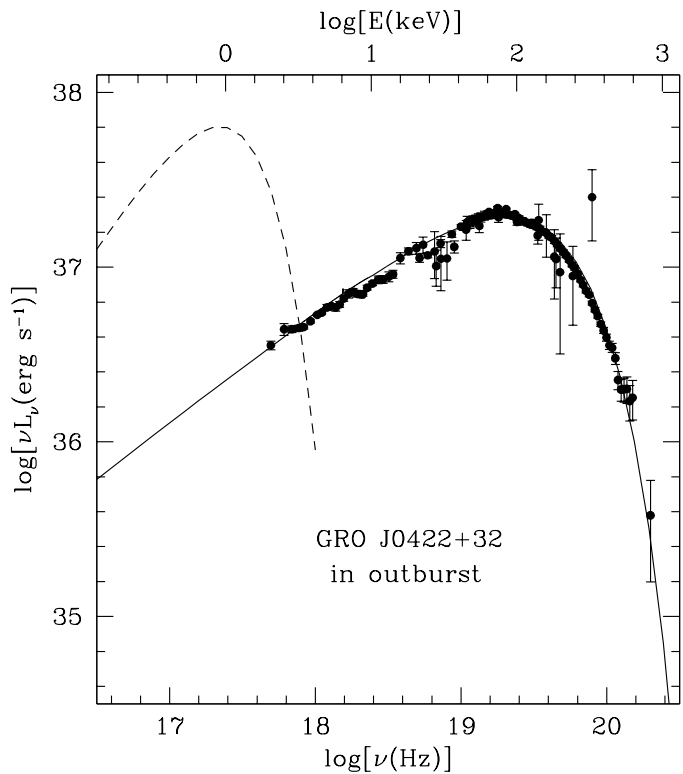}
}
\caption{An ADAF model of J0442+32 (solid line) in the low state, compared
with the observational data (dots and errorbars).  The dashed line
shows a thin disk model at the same accretion rate, $\dot m = 0.1$.}
\label{0442}
\end{figure}

\begin{enumerate}
\item Quiescent state: This lowest luminosity state has $\dot{m} \lsim
10^{-2}$ and is discussed above (\S4.1.1).  Due to the low accretion
rate, Comptonization is weak, and the X-ray flux is much lower than
the optical flux.  The radiative efficiency is very low and the
systems are extremely dim (cf. Figs. \ref{spectra} and \ref{lum}).
\item Low state: For $\dot{m}$ above $10^{-2}$ and up to
$\sim10^{-1}$, the geometry of the accretion flow is similar to that
of the quiescent state, but the luminosity and radiative efficiency
are larger (and increase rapidly with $\dot m$).  A low state spectrum
of GRO J0442+32 is shown in figure (\ref{0442}).  Comptonization
becomes increasingly important, giving rise to a very hard spectrum
which peaks around 100 keV.
\item Intermediate state: At still higher accretion rates, $\dot{m}$
approaches $\dot{m}_{\rm crit} \sim 0.1$, the ADAF progressively
shrinks in size, the transition radius decreases, and the X-ray
spectrum changes continuously from hard to soft.  This occurs at
roughly constant bolometric luminosity.  In this state, the thin disk
becomes radiatively comparable to, or even brighter than, the ADAF.
\item High state: At still higher accretion rates, $\dot{m} >
\dot{m}_{\rm crit}$, the ADAF cannot exist as an independent entity at
any radius, and the thin disk comes all the way down to the last
stable orbit; there is, however, a weak corona above the thin disk
which is modeled as an ADAF (with a coronal ${\dot m} \lsim {\dot
m}_{\rm crit}$). A characteristic high state spectrum resembles a
standard thin disk spectrum, with a power law tail due to the corona.
\end{enumerate}
This model accounts convincingly for the characteristic spectral state
variations, from quiescence to the high state, seen in black hole
X-ray binaries.  

\begin{figure}
\centerline{
\psfig{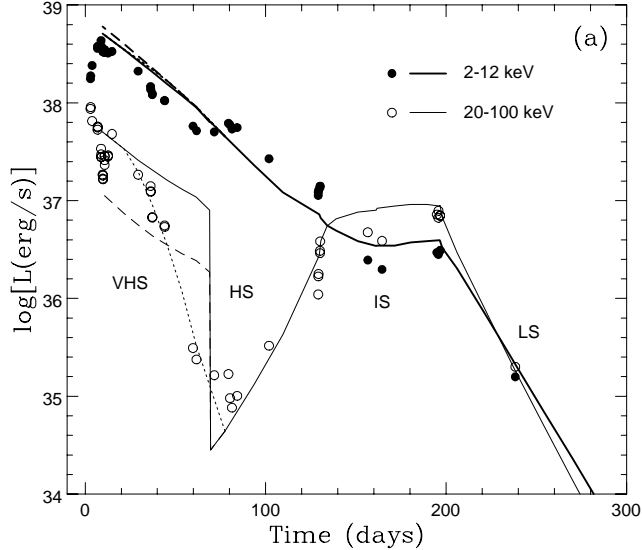}
}
\caption{Soft and hard X-ray light curves of Nova Muscae 1991.  Filled
and open circles are data from Ebisawa et al. (1994) corresponding to
the 2--12 keV and 20--100 keV bands, respectively.  The heavy and thin
lines are the model predictions (from Esin et al. 1997).  The symbols
VHS, HS, IS, LS correspond to the very high state, high state,
intermediate state and low state, respectively.  In the very high
state, the solid, dotted, and dashed lines correspond to different
prescriptions for viscous dissipation in the corona (see Esin et
al. 1997 for details).}
\label{esinfig12}
\end{figure}

Esin et al. (1997; 1998) applied their model to the 1991 outburst of
the SXT Nova Muscae and to the high-low/low-high state transition of
Cyg X-1.  Nova Muscae went into outburst during the fall of 1991, and
was extensively studied in the optical and X--ray bands (Ebisawa et
al. 1994; see also Brandt et al. 1992; Lund 1993; Goldwurm et
al. 1992; Gilfanov et al. 1993).  Its luminosity evolution is
summarized in figure (\ref{esinfig12}).  The model light curves agree
quite well with the data.

\begin{figure}
\centerline{
\psfig{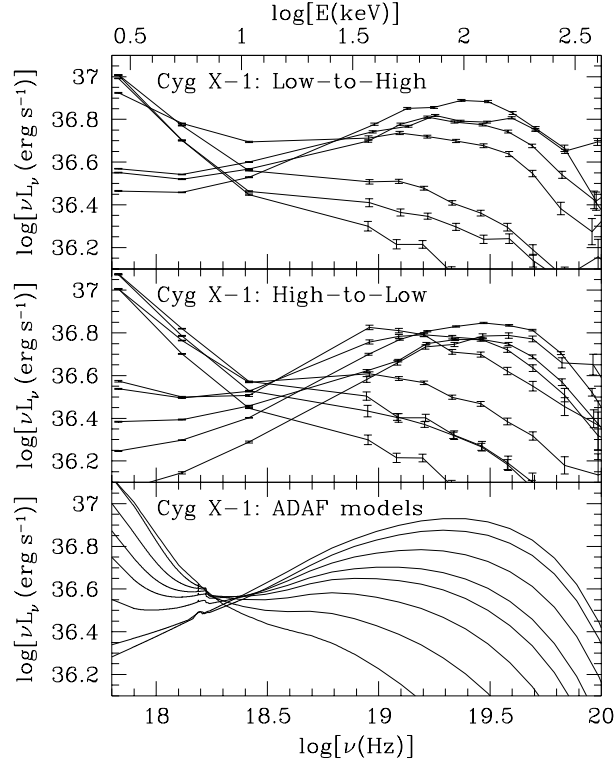}
}
\caption{The RXTE (1.3-12 keV) and BATSE (20-600 keV) spectra of Cyg
X-1 during the 1996 low-high (upper panel) and high-low (middle panel)
state transitions.  The bottom panel shows a sequence of ADAF + thin
disk models which are in good agreement with the observations.}
\label{cygx1}
\end{figure}

Figure (\ref{cygx1}) shows the broadband simultaneous RXTE (1.3-12
keV) and BATSE (20-600 keV) spectra of Cyg X-1 observed during the
1996 low-high (upper panel) and high-low (middle panel) state
transitions.  The bottom panel shows a sequence of ADAF + thin disk
models, which are in good agreement with the observations.  In
particular, the model reproduces the range of photon indices seen in
the data, the anti-correlation between the soft and hard X-ray flux,
the ``pivoting'' around 10 keV, and the nearly constant bolometric
luminosity throughout the transition.

One important result, highlighted by Narayan (1996), and confirmed by
the more detailed calculations of Esin et al. (1997), is that the
application of ADAFs to luminous black hole X-ray binaries requires a
fairly large value of $\alpha\sim0.25$.  (The preliminary work of
Narayan 1996 actually suggested $\alpha\sim1$, but this was revised in
Esin et al. 1997.)  If $\alpha$ is much smaller, then the ADAF
solution cuts off at a very low mass accretion rate (recall that $\dot
m_{\rm crit}\sim\alpha^2$), and the observed luminosities of low state
systems cannot be explained.  As discussed in \S3.1, $\alpha$ will
likely be large if the accreting gas has magnetic fields of
equipartition strength.

Some black hole X-ray binaries, such as Nova Muscae 1991, exhibit an
even higher luminosity state, called the very high state, which is
significantly harder than the high state.  This does not fit readily
into the thin disk + ADAF paradigm.  Esin et al. (1997) showed that if
there is enhanced viscous dissipation of energy in the corona, as
suggested by Haardt \& Maraschi (1991, 1993) for modeling coronae in
AGN, some properties of the very high state can be understood.  Esin
et al.'s model of the very high state is, however, quite speculative,
and does not fit the observations particularly well.

\subsection{Applications to Galactic Nuclei}

\subsubsection{Sgr A*}

\begin{figure}
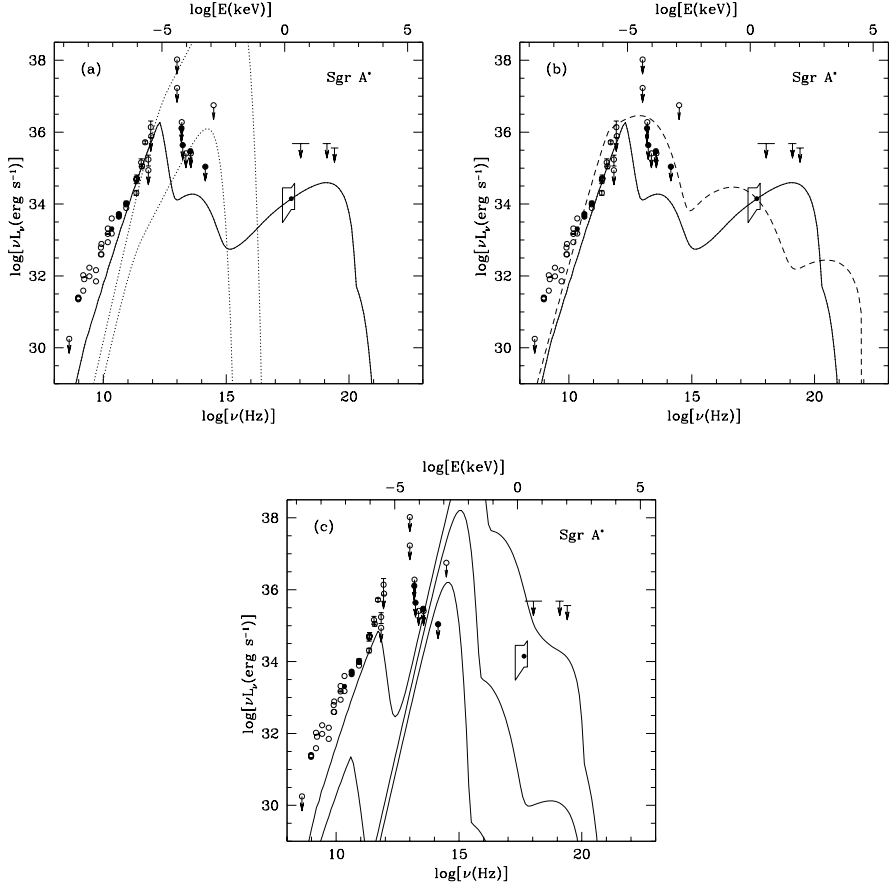

\begin{center}
\begin{minipage}[t]{0.45\hsize}
\begin{displaymath}
\psfig{figure=nfig15a.epsi,height=2.3in}
\end{displaymath}
\end{minipage}
\begin{minipage}[t]{0.45\hsize}
\begin{displaymath}
\psfig{figure=nfig15b.epsi,height=2.3in}
\end{displaymath}
\end{minipage}
\begin{minipage}[t]{0.45\hsize}
\begin{displaymath}
\psfig{figure=nfig15c.epsi,height=2.3in}
\end{displaymath}
\end{minipage}
\end{center}
\caption{(a) Spectrum of a two temperature ADAF model of Sgr A* (solid
line).  The mass accretion rate inferred from this model is $\dot m =
1.3 \times 10^{-4}$, in agreement with an independent observational
determination.  Dotted lines show the spectra of thin accretion disks,
at the same accretion rate (upper) and at $\dot m = 10^{-8}$ (lower).
(b) The dashed line is a one temperature ADAF model of Sgr A* at $\dot
m = 3 \times 10^{-7}$ (adjusted to fit the x-ray flux).  The solid
line is the standard two temperature model.  (c) Spectra of ADAF
models of Sgr A* where the central star is taken to have a hard
surface at $r = 3$ and the advected energy is assumed to be reradiated
from the surface as a blackbody.  The three spectra correspond to
$\dot m = 10^{-4}, 10^{-6},$ and $10^{-8}$ (from top to bottom).  All
three models violate the infrared limits.}
\label{sgr}
\end{figure}
 
Dynamical measurements of stellar velocities within the central 0.1 pc
of the Galactic Center indicate a dark mass with $M \sim (2.5 \pm 0.4)
\times 10^6M_{\odot}$ (Haller et al. 1996; Eckart \& Genzel 1997).
This is believed to be the mass of the supermassive black hole in Sgr
A$^*$.  Observations of stellar winds indicate that the expected
accretion rate is $6\times 10^{-6}M_{\odot}$ yr$^{-1} \leq \dot{M}
\leq 2 \times 10^{-4}M_{\odot}$ yr$^{-1}$ (Genzel et al. 1994; Melia
1992), which corresponds to $10^{-4} \leq \dot{m} \leq 3\times
10^{-3}$. Using a nominal radiative efficiency of 10\% these accretion
rates imply an accretion luminosity ($\sim 0.1 \dot{M} c^2$) between
$\sim 10^{40} $erg s$^{-1}$ and $\sim 10^{42}$erg s$^{-1}$.
Observations in the radio to $\gamma$--rays, however, seem to indicate
a bolometric luminosity of less than $10^{37}$erg s$^{-1}$.  This
extremely low luminosity has been used to argue against a supermassive
black hole in Sgr A$^*$ (Mastichiadis \& Ozernoy 1994; Goldwurm et
al. 1994).

A standard thin disk with the above accretion rate gives rise to a
black body spectrum which peaks in the near infrared (Frank et
al. 1992).  Menten et al. (1997), however, have obtained a strong upper
limit on the infrared emission from the Galactic center, and find it
to be below $\sim 10^{34}$erg s$^{-1}$.  This effectively rules out
any thin disk model of the Galactic Center.

All of the apparently contradictory observations of Sgr A* appear to
be naturally accounted for by an optically thin, two temperature, ADAF
model.  Rees (1982) first suggested that Sgr A* may be advecting a
significant amount of energy and Narayan et al. (1995) provided the
first spectral model; more detailed recent models (Manmoto et
al. 1997; Narayan et al. 1998) confirm that an ADAF model can explain
the observations quite well.  The solid line in Figure (\ref{sgr}a)
shows the best fit Narayan et al. (1998) model, while the dotted lines
correspond to thin disk models, which are easily ruled out by the
data.  The mass of the central black hole is fixed at its dynamically
measured value of $2.5 \times 10^6 M_{\odot}$ and the accretion rate
is varied to fit the X--ray flux.  The resulting accretion rate is
$\dot{m} \sim 1.3 \times 10^{-4}$, which is consistent with the mass
accretion rate estimates from the observations of stellar winds.

The resulting model naturally reproduces the observed spectrum in
other wavebands.  Synchrotron radiation from the ADAF produces the
radio spectrum which cuts off sharply in the sub--mm; the inverse
Compton spectrum is consistent with the stringent upper limit on the
NIR flux given by Menten et al. (1997); finally, bremsstrahlung
radiation is responsible for the X--ray flux which extends up to a few
hundred keV.  There is, however, a problem at low radio frequencies,
$\lsim 10^{10}$ Hz, where the model is well below the observed flux.
In addition, the $\gamma$--ray spectrum from the ADAF is lower than
the observed flux by nearly an order of magnitude (not shown in Figure
\ref{sgr}; see Figure 1 of Narayan et al. (1998)).  The latter is not
considered to be a serious problem since it is unclear whether the
observed $\gamma$--rays are in fact from a point or a diffuse source
at the Galactic Center.

Sgr A* is one of the few observed systems for which the luminosity is
low enough that a one temperature ADAF model can be constructed
(recall from \S3.2.2 that ${\dot m}_{\rm crit} \sim 10^{-6}$ for a one
temperature ADAF so that a source must have a luminosity $\lsim
10^{-6} L_{\rm edd}$ for a one temperature ADAF model to be possible).
This model (dashed line in Figure \ref{sgr}b) is, however, ruled out
by the data.

An important feature of the two temperature ADAF model of Sgr A* is
that the observed low luminosity is explained as a natural consequence
of the advection of energy in the flow, rather than as a very low
accretion rate (the radiative efficiency is very low, $\sim 5 \times
10^{-6})$.  The model will not work if the central object has a hard
surface and reradiates the advected energy, as demonstrated in Figure
(\ref{sgr}c).  Therefore, the success of the ADAF model implies that
Sgr A$^*$ is a black hole with an event horizon (\S4.3).

\subsubsection{NGC 4258}

\begin{figure}
\centerline{
\psfig{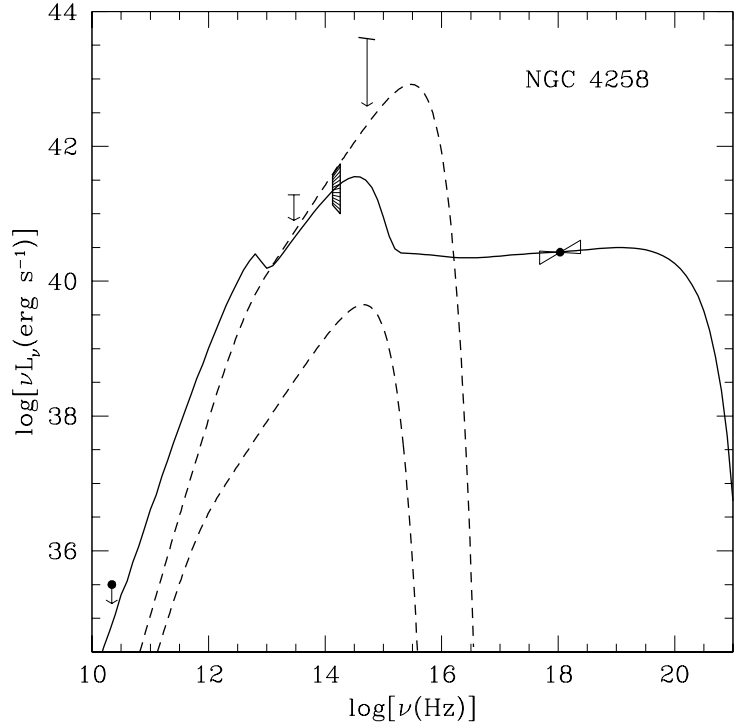}
}
\caption{Spectrum of an ADAF model of NGC 4258 (solid line), at an
accretion rate of $\dot m = 9 \times 10^{-3}$ with a transition
radius of $r_{tr} = 30$.  Dotted lines show the spectra of thin
accretion disks at accretion rates of $\dot m = 4 \times 10^{-3}$
(upper, adjusted to fit the infrared points) and $\dot m = 10^{-5}$
(lower).}
\label{4258}
\end{figure}
 
The mass of the central black hole in the AGN NGC 4258 has been
measured to be $ 3.6 \times 10^7 M_{\odot}$ (Miyoshi et al. 1997).
Observations of water masers indicate the presence of a thin disk, at
least at large radii.  Highlighting the fact that the observed
optical/UV and X-ray luminosities are significantly sub-Eddington
($\sim 10^{ -4}$ and $\sim 10^{-5}$, respectively), Lasota et
al. (1996) proposed that the accretion in NGC 4258 at small radii
proceeds through an ADAF.  In Lasota et al.'s original model, the
transition radius was a free parameter, but since then new infrared
observations have been made (Chary \& Becklin 1997) that constrain the
transition radius to be $r_{tr} \sim 30$.  The outer thin disk then
accounts for the newly observed infrared emission, and the refined
model (see Fig. \ref{4258}, based on Gammie, Narayan \& Blandford
1998) is also in agreement with a revised (smaller) upper limit on the
radio flux from NGC 4258 (Herrnstein et al. 1998).  Maoz \& McKee
(1997) and Kumar (1997) find, via quite independent arguments, an
accretion rate for NGC 4258 in agreement with the ADAF model ($\dot m
\sim 0.01$).  Neufeld \& Maloney (1995) estimate a much lower
$\dot{m}<10^{-5}$ in the outer maser disk via a model of the maser
emission.  The accretion rate close to the black hole cannot be this
low; not even an efficiently radiating thin disk can produce the
observed infrared and X-ray radiation with such an $\dot m$ (the lower
dotted line in Fig. \ref{4258}).

\subsubsection{Other Low Luminosity Galactic Nuclei}

ADAFs have also been used to model a number of other nearby low
luminosity galactic nuclei, e.g., M87 (Reynolds et al. 1996) and M60
(Di Matteo \& Fabian 1997a) in the Virgo cluster.  These and similar
elliptical galaxies are believed to have contained quasars with black
hole masses $\sim 10^8-10^9M_{\odot}$ at high redshift (Soltan 1982;
Fabian \& Canizares 1988; Fabian \& Rees 1995).  In the case of M87
and several other galaxies there is independent evidence for such dark
mass concentrations in their centers (Ford et al. 1997).  The unusual
dimness of these galactic nuclei is, however, a problem.

Fabian \& Canizares (1988) considered six bright nearby ellipticals
and, using X--ray gas profiles in the central arcsecond regions,
estimated the accretion rates onto the central black holes.  For a
standard radiative efficiency of $0.1$, their estimated accretion
rates yielded luminosities which were substantially larger than the
observed luminosities.  Thus these galactic nuclei appeared to be very
underluminous.  Fabian \& Canizares (1988) highlighted this problem,
calling it the dead quasar problem.  As suggested by Fabian \& Rees
(1995), and confirmed by the detailed calculations of Mahadevan
(1997), the problem is naturally resolved if the galactic nuclei are
presently accreting via an ADAF, rather than a thin disk.  ADAFs are
naturally underluminous and have low radiative efficiencies.
Similarly, Lasota et al. (1996b) suggested that perhaps all LINERs (of
which NGC 4258 is an example) have ADAFs.

\subsubsection{AGN Statistics}

Quasars first appear at a redshift $z \sim 5$ and their numbers
increase with decreasing $z$; below a redshift $\sim 2$, however, the
number of quasars decreases rapidly. Quasars are essentially
non--existent at the present epoch, $z\sim 0$.  In the standard AGN
paradigm, all quasars are assumed to be powered by accretion onto
supermassive black holes. Yi (1996) has suggested (following Fabian \&
Rees 1995) that quasars may switch from accretion via a thin disk to
accretion via an ADAF at $1 \lsim z \lsim 2$; this provides a natural
explanation for the decrease in quasar number counts at small $z$
since ADAFs are significantly less luminous and thus much more
difficult to detect.  What accounts, however, for the change in the
accretion mechanism at $z \sim 2$?  Yi assumes that at large $z$,
quasars accrete at $\dot{m} \sim 1$, and so the accretion must be via
a thin disk.  As the quasar evolves, however, two processes lead to
decreasing $\dot{m}$: (1) A decrease in the fuel supply, and (2) an
increase in the mass of the accreting black hole; since $\dot{m}
\propto \dot{M}/M$, both of these cause $\dot m$ to decrease and lead
to a critical redshift below which $\dot{m} \lsim \dot{m}_{\rm crit}$;
at this point ($z \sim 2$), the accretion flow switches to an ADAF.

\subsubsection{The X-Ray Background}

Di Matteo \& Fabian (1997b; see also Yi \& Boughn 1998) argue that
ADAFs can be used to model the diffuse X--ray background.  The
spectrum of the diffuse XRB resembles thermal bremsstrahlung in the
3--60 keV range, and has a rollover at $\sim 30$ keV. The X--ray
background is thought to arise from many discrete sources.  ADAFs are
a natural candidate since (1) they intrinsically produce
bremsstrahlung, (2) the electron temperature is in the right range,
$\sim 10^9$K, to account for the observed cutoff, and (3) the electron
temperature is very insensitive to the parameters of the model.
Assuming a modest spectral evolution with redshift, Di Matteo \&
Fabian (1997b) were able to reproduce the X--ray background fairly
well.  Since moderately high accretion rates are required to account
for the observed flux, however, Comptonization contributes to the hard
X--ray spectrum of the ADAFs.  Di Matteo \& Fabian avoid this problem
by invoking a high degree of clumpiness in the gas so that
bremsstrahlung emission dominates over Compton scattering, but this is
somewhat ad hoc.

\subsection{Evidence for the Black Hole Event Horizon}

\begin{figure}
\centerline{
\psfig{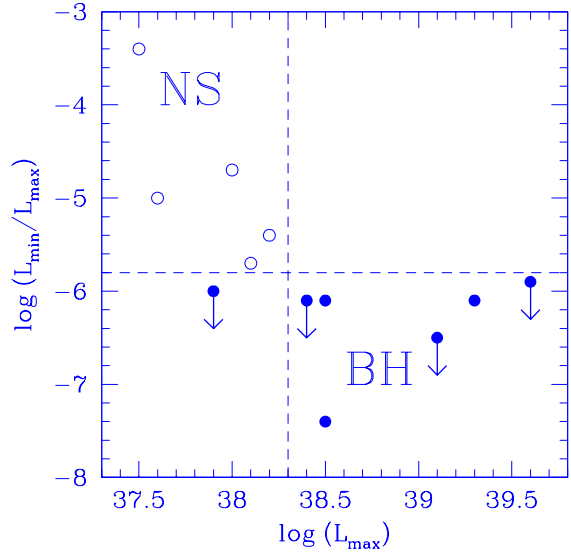}
}
\caption{A comparison between black hole (BH, filled circles) and
neutron star (NS, open circles) SXT luminosity variations (from
Narayan et al. 1997c and Garcia et al. 1998).  The ratio of the
quiescent luminosity to the peak outburst luminosity is systematically
smaller for BH systems than for NS systems. This indicates the
presence of event horizons at the center of BH candidate systems.}
\label{sxt}
\end{figure}

A unique feature of the ADAF applications described above is that the
models require the existence of an event horizon in the central object
(Narayan et al. 1996; Narayan et al. 1997bc).  Since the protons store
most of the viscous energy as thermal energy as they fall onto the
central object, the luminosity from the ADAF is much less than the
viscously generated energy ($\sim 0.1 {\dot M} c^2$).  If the central
object had a hard surface, this thermal energy would be reprocessed
and reradiated, resulting in a net luminosity from the object $\sim
0.1 {\dot M} c^2$; the spectrum would also differ significantly from
that of a pure ADAF (since it would be dominated by the reprocessed
radiation; see Figure \ref{sgr} as an example).  If the central object
is a black hole, however, the advected thermal energy is carried into
the black hole through the event horizon, and is ``lost'' to the
outside observer.  The success of ADAF models of the various black
hole candidates discussed in the previous sections {\em without}
reprocessed radiation therefore indicates the presence of an event
horizon in all of these objects.

Narayan et al. (1997c; see also Yi et al. 1996) highlighted the
distinction between black holes and compact stars with surfaces in a
different way by comparing the luminosities of black hole and neutron
star soft X--ray transients in outburst and quiescence.  In outburst,
the accretion rates in both neutron star and black hole systems are
high, ${\dot m} \sim 1 \gsim {\dot m}_{\rm crit}$, and the accretion
occurs via a standard thin disk.  The observed outburst luminosities,
$L_{max}$, should therefore be proportional to $\dot{M} \propto m$.
This is consistent with the observation that black hole SXTs in
outburst have a larger $L_{max}$ than neutron star SXTs, as expected
because of their larger mass.

In quiescence, the accretion rates are low, $\dot m\ll\dot m_{\rm
crit}$, and the accretion occurs via an ADAF.  Since neutron stars
have hard surfaces, however, the advected energy is ultimately
reradiated, and the luminosity is still proportional to $\dot{m}$.
For an ADAF around a black hole, on the other hand, the advected
energy is not reradiated, but is deposited into the black hole.  The
quiescent luminosities, $L_{min}$, from black hole candidates should
therefore be substantially less, roughly proportional to ${\dot m}^2$
(see Fig. \ref{lum}).  Black hole SXTs should thus experience more
substantial luminosity changes from quiescence to outburst than
neutron star SXTs. As figure (\ref{sxt}) shows, this is exactly what
is observed.  This provides additional evidence for the existence of
event horizons in black hole candidates.

\section{Conclusion}

The two-temperature ADAF model provides a consistent framework for
understanding the dynamics and spectra of black hole accretion flows
at low mass accretion rates, $\dot m\lsim0.1$.  By modeling coronae as
ADAFs, the model is also being extended successfully to systems with
$\dot m\gsim0.1$, though not yet to systems close to the Eddington
limit ($\dot m\sim1$).

The field is young and there are many open questions.  Perhaps the
most fundamental issue is whether the assumptions underlying the
two-temperature ADAF model are valid (\S3.1).  Another major question,
where much work needs to be done, is the physics underlying the
transition radius (\S3.3).  If this is understood, and the dynamics
and thermal structure of the corona can be modeled well, it will be
possible to calculate $r_{tr}(\dot m)$ reliably; this will be an
enormous improvement over the present work.  Improvements are
necessary in the modeling techniques as well.  Fully relativistic
computations of the coupled dynamics and radiative transfer will be
welcome.  One needs to include time-dependence in the models in order
to understand the complex variations of spectra with time, especially
in X-ray binaries such as the SXTs.  Ultimately, we expect that
time-dependent two-dimensional simulations will elevate the ADAF model
to a fully quantitative tool, though this is probably many years away.

Two wider issues should be highlighted.  First, the high-$\dot m$,
optically thick, ADAF (corresponding to radiation trapping) has not
been explored at the level of detail of the low $\dot m$,
two-temperature, ADAF.  More quantitative models, especially with
respect to their spectral properties, would be worthwhile.  It is
possible that some of the presently most puzzling sources (e.g. SS433)
correspond to this branch of ADAFs.  The other issue has to do with
jets and outflows.  There are tantalizing hints that ADAFs may be
particularly efficient at producing outflows (Narayan \& Yi 1994,
1995a), but the connection has not been developed in any detail.

We thank Kristen Menou for comments on the manuscript.  This work was
supported in part by NSF Grant AST 9423209 and NASA Grant 5-2837.  EQ
was supported in part by an NSF Graduate Research Fellowship.

\end{document}